\documentclass{emulateapj}
\slugcomment{draft v1}

\shorttitle{Synchrotron Radiation from Pulsars}
\shortauthors{Kisaka \& Tanaka}

\begin{document}


\title{Efficiency of Synchrotron Radiation from Rotation-Powered Pulsars}


\author{Shota Kisaka\altaffilmark{1,2}}
\email{kisaka@phys.aoyama.ac.jp}
\author{Shuta J. Tanaka\altaffilmark{3}}
\email{sjtanaka@center.konan-u.ac.jp}


\altaffiltext{1}{Department of Physics and Mathematics, Aoyama Gakuin University, Sagamihara, Kanagawa, 252-5258, Japan}
\altaffiltext{2}{JSPS Research Fellow}
\altaffiltext{3}{Department of Physics, Konan University, Kobe, Hyogo, 658-8501, Japan}


\begin{abstract}

Synchrotron radiation is widely considered as the origin of the pulsed non-thermal emissions 
from rotation-powered pulsars in optical and X-ray bands. 
In this paper, we study the synchrotron radiation emitted by the created electron and 
positron pairs in the pulsar magnetosphere 
to constrain on the energy conversion efficiency from the Poynting flux to 
the particle energy flux.  
We model two pair creation processes, two-photon collision which efficiently works 
in young $\gamma$-ray pulsars ($\lesssim10^6$ yr), and magnetic pair creation
which is the dominant process to supply pairs in old pulsars ($\gtrsim10^6$ yr). 
Using the analytical model, we derive the maximum synchrotron luminosity as a function of the energy conversion efficiency. 
From the comparison with observations, 
we find that the energy conversion efficiency to the accelerated particles should be 
an order of unity in the magnetosphere, 
even though we make a number of the optimistic assumptions to enlarge the synchrotron luminosity. 
In order to explain the luminosity of the non-thermal X-ray/optical emission 
from pulsars with low spin-down luminosity $L_{\rm sd}\lesssim10^{34}$ erg s$^{-1}$, 
non-dipole magnetic field components should be dominant at the emission region.
For the $\gamma$-ray pulsars with $L_{\rm sd}\lesssim10^{35}$ erg s$^{-1}$, observed 
$\gamma$-ray to X-ray and optical flux ratios are much higher than the flux ratio between curvature  
and the synchrotron radiations. 
We discuss some possibilities such as the coexistence of multiple accelerators 
in the magnetosphere as suggested from the recent numerical simulation results. 
The obtained maximum luminosity would be useful to select observational targets 
in X-ray and optical bands.

\end{abstract}


\keywords{ ---  --- }



\section{INTRODUCTION}
\label{sec:introduction}

Rotation-powered pulsars are capable of producing pulsed emissions with wide energy bands. 
In some regions where the charge density is depleted from the Goldreich-Julian (GJ) 
charge density \citep{GJ69}, particles are accelerated and emit high-energy photons.
High-energy photons convert to electron and positron pairs which screen the accelerating electric field. 
Then, the particle acceleration and emission regions are spatially limited, which is of the pulsed emission 
\citep[e.g., ][]{S71, AS79, CHR86}. 
During the processes, a part of the rotation-energy flux of a neutron star $L_{\rm sd}$ converts to 
the particle kinetic energy flux. 
The primary particles which are significantly accelerated by the electric field emit the curvature radiation, 
and the created secondary (and higher generation) pairs with non-zero pitch angle emit synchrotron radiation. 
Thus, observations of pulsed emissions provide the location of particle acceleration and particle creation, 
and the efficiency of the energy conversion to the particle energy.

Pulsed emission in GeV $\gamma$-ray band is an important tool for probing the particle acceleration 
in the magnetosphere.
GeV $\gamma$-ray emission is detected in young pulsars with characteristic age $\tau_{\rm c}\lesssim10^6$ yr, 
except for millisecond pulsars \citep{Ab13}.
The emission mechanism is considered as the curvature radiation from the primary particles. 
Since the GeV spectrum observed by {\it Fermi} show that the cutoffs are more gradual than exponential 
\citep{Ab09, Ab13}, 
the emission comes from the outer region of the magnetosphere such as the outer gap model 
\citep[e.g., ][]{CHR86}. 
The observed cutoff energy is distributed around $\sim$ GeV \citep{Ab13}, 
which provides the energy of the accelerated particles. 
The observed $\gamma$-ray luminosity roughly seems to follow the trend $L_{\gamma}\propto L_{\rm sd}^{1/2}$ 
\citep{Ab13}. 
The trend is expected in models that the $\gamma$-ray luminosity is 
simply proportional to the GJ current \citep{H81}. 

Although the $\gamma$-ray observations provide the constraints on the energy conversion efficiency 
and the location of the particle acceleration, 
the information could be inadequate to understand the particle acceleration for the whole population of 
rotation-powered pulsars.
Since GeV $\gamma$-ray emissions are detected from only young pulsars, 
the efficiency of energy conversion is less constrained for older, non-$\gamma$-ray pulsars. 
It is difficult to obtain the reliable value of the $\gamma$-ray luminosity of the $\gamma$-ray 
pulsars that are not detected in radio because of the uncertain distance from the observer.
In addition, the energy conversion efficiency may be underestimated by only $\gamma$-ray observations
because of the pair creation process. 
For example, the inner accelerator could also work to supply huge number of pairs 
required in the observations of pulsar wind nebulae \citep[e.g., ][]{TT10, TT11, TT13}. 
The $\gamma$-ray photons from the inner region are significantly absorbed via magnetic (B$\gamma$) 
pair creation process, 
so that total energy flux radiated in $\gamma$-ray is larger than that estimated 
from the $\gamma$-ray observations.

The non-thermal pulsed emissions are also detected in lower energy bands such as X-ray and optical bands. 
The non-thermal X-ray luminosity seems to follow the relation $L_{\rm X}\propto L_{\rm sd}^{a}$ with $a\sim0.9-1.5$
\citep[e.g., ][]{SW88, BT97, Pos02, B09}, 
although scatter about the relation is significant \citep{Kar12, Sh16}.
Critical lines below which all the data locates on $L_{\rm X}$ versus $L_{\rm sd}$ plane 
are empirically suggested \citep{Pos02, Kar12}, 
although the physical mechanism of the restriction of X-ray luminosity is unknown. 
The non-thermal optical luminosity also seems to depend on the spin-down luminosity 
in a manner similar to the X-ray luminosity 
\citep[e.g., ][]{ZSK06}. 
The emission mechanism for both X-ray and optical bands is widely believed 
as the synchrotron radiation from the secondary 
(and higher generation) particles \citep[e.g., ][]{R96, ZC97, HSDF08, TCS08}. 

The X-ray and optical emissions, and the combination with $\gamma$-ray observations provide 
valuable information about the particle acceleration. 
In contrast to GeV $\gamma$-ray, the X-ray and optical emissions are detected even from old pulsars 
whose ages are up to $\sim10^8$ yr \citep{Pos+12}. 
A photon whose energy is much less than a electron rest mass energy 
is not absorbed via the magnetic pair creation process. 
Then, we can detect the X-ray and optical emissions even if their emission regions reside 
at the inner region of the magnetosphere.
These emissions provide the energy flux of secondary particles created in the magnetosphere 
for almost all pulsars. 
Since $\gamma$-ray, X-ray and optical emitting particles are related through a pair cascade process,
if the $\gamma$-ray emission is detected, their flux ratios which do not depend on the distance 
are useful to understand the pair cascade process in the magnetosphere. 

One of the reasons for the investigation of the energy conversion for old pulsars and the comparison with 
that for young pulsars is that some of old pulsars have intensity and pulse profile modulations
such as nulling and mode changing in radio observations \citep[e.g., ][]{B70a, B70b}.
The fraction of these modulated pulsars significantly increases from their characteristic age 
$\tau_{\rm c}\gtrsim10^6$ yr \citep[e.g., ][]{WMJ07}, 
although it is unknown whether the characteristic age is a control parameter 
for the modulation phenomena or not. 
Recently, the correlation between the state changes of radio emission 
and the change of braking properties have been discovered \citep{K06, Lyne10}. 
This correlation suggests that the modulation phenomena are linked 
and caused by the change of the magnetospheric state. 
Moreover, the synchronous radio and X-ray switching between two modes 
was also reported \citep{Her13, Mer+13, Mer+16}. 
Therefore, the differences of the location of the X-ray emission region 
and the energy conversion efficiency between young and old pulsars 
give hints to understand such modulation phenomena. 

Recently, enegy dissipation in the magnetosphere has been investigated by numerical simulations 
such as dissipative magnetohydrodynamic approaches \citep{LST12, Kal+12, KHK14} and global Particle-in-Cell simulations 
\citep{PS14, CB14, B15, PSC15, PCTS15, CPPS15, CPS16}. 
Their results indicate that most particle acceleration takes place in the current sheet 
close behind the light cylinder \citep{CB16}. 
The X-ray and optical synchrotron radiations from pairs in the current sheet have been discussed \citep[e.g., ][]{L96}. 
Hence, the X-ray and optical emissions would provide valuable constraints on the current sheet emitting scenario. 

In this paper, using the analytical model, 
we calculate the luminosity of the synchrotron radiation 
as the emission mechanism of the non-thermal X-ray and optical bands. 
We parameterize the efficiency of energy conversion from the spin-down luminosity 
to the energy flux of primary particles.
We are interested in the efficiency of the energy conversion and the emission region 
in whole population including $\gamma$-ray detected pulsars ($\tau_{\rm c}\lesssim10^6$ yr) 
and old pulsars ($\tau_{\rm c}\gtrsim10^6$ yr). 
In $\gamma$-ray pulsars, 
pair creation occurs at the outer magnetosphere via two-photon collision ($\gamma\gamma$).
As pulsar gets old, it is suggested that $\gamma\gamma$ pair creation becomes ineffective 
at the outer magnetosphere \citep{WH11, KT14}.
Thus, we consider two pair creation processes, B$\gamma$ and $\gamma\gamma$ pair creations. 
In previous study \citep{KT14}, we assume that the energy of the primary particles equals to 
the maximum potential drop across the polar cap. 
Using this assumption, we avoid assuming the value of accelerating electric field which is highly uncertain.
For young pulsars, the characteristic energy of the curvature radiation from such primary particles 
is much higher than the observed cutoff energy. 
Here, we use the typical value of the observed cutoff energy in GeV $\gamma$-ray band
to derive the energy of primary particles in $\gamma\gamma$ pair creation case. 
In B$\gamma$ pair creation case, we use the maximum potential drop across the polar cap to 
derive the energy of primary particles. 
Note that in this case, the results do not depend on the energy of primary particles in our model. 
In section \ref{sec:model}, we describe our model for the radiative transfer and the luminosity of the synchrotron radiation. 
In section \ref{sec:results}, we provide the luminosity of the synchrotron radiation as a function of the efficiency of 
the energy conversion for each pair creation case. 
We also provide the allowed range of the emission region of the synchrotron radiation in X-ray and optical bands.
Discussion is presented in section \ref{sec:discussion}. 

\section{MODEL}
\label{sec:model}

\subsection{Assumptions}
\label{sec:assumptions}

 \begin{figure*}
  \begin{center}
   \includegraphics[width=120mm]{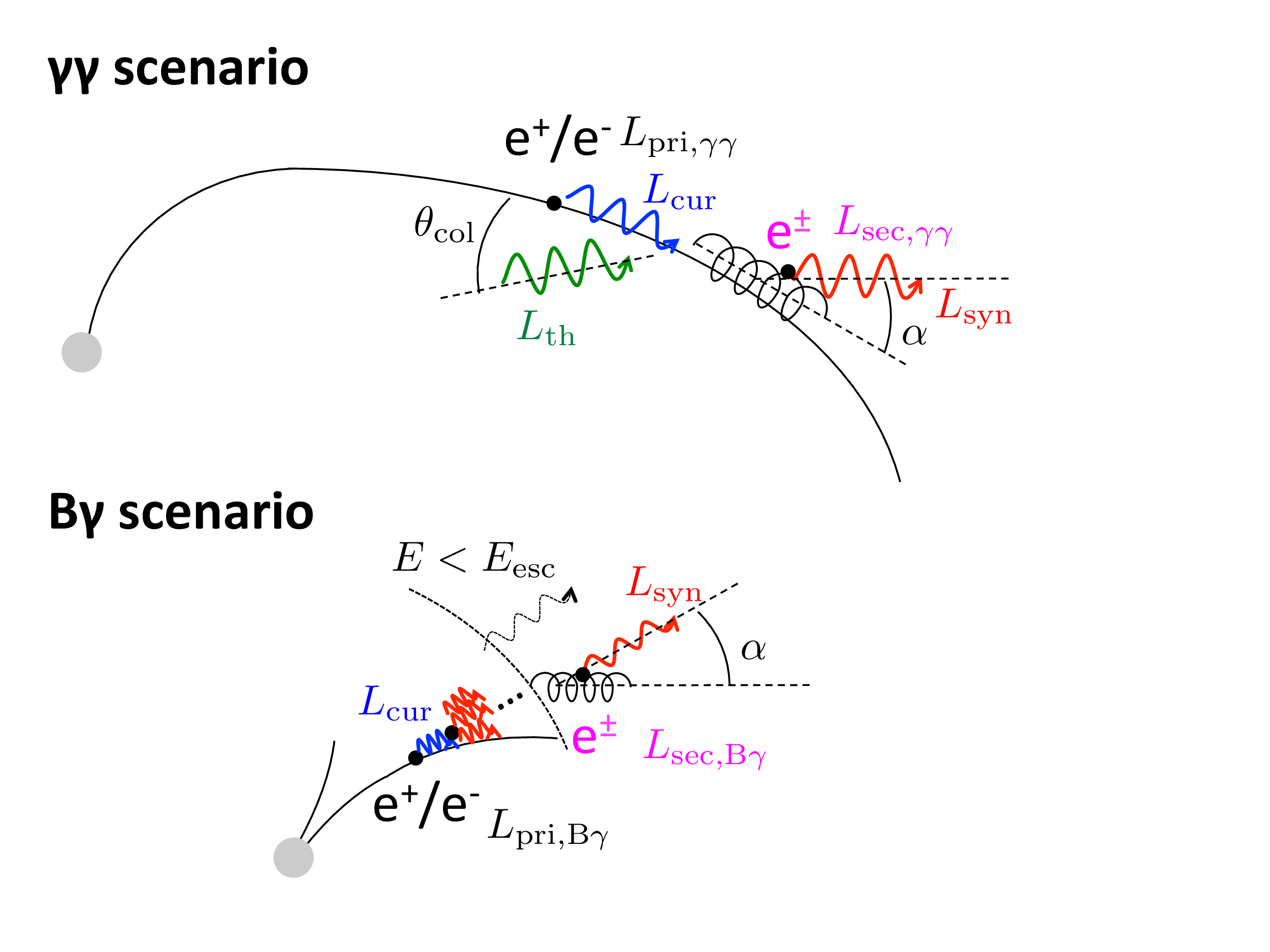}
   \caption{Schematic pictures of our model for $\gamma\gamma$ (upper panel) and B$\gamma$ scenarios (lower panel). 
Grey circles are neutron stars, and thin solid lines denote the magnetic field lines. 
Blue, green, and red arrows denote the photons of curvature radiation from primary particle, 
thermal radiation from heated polar cap, and synchrotron radiation from secondary (and higher generation) pairs. 
The definitions of the energy fluxes for the primary and secondary particles in $\gamma\gamma$ scenario are 
$L_{\rm pri, \gamma\gamma}\equiv\min\{\gamma_{\rm p,E},\gamma_{\rm p,max}\}m_{\rm e}c^2\dot{N}_{\rm p}$ and 
$L_{\rm sec,\gamma\gamma}\equiv2\gamma_{\rm s,syn}m_{\rm e}c^2\dot{N}_{\rm p}N_{\gamma}\tau_{\gamma\gamma}$, respectively.
On the other hand, in B$\gamma$ scenario, 
the energy fluxes of the primary and secondary particles are 
$L_{\rm pri, B\gamma}\equiv\gamma_{\rm p,max}m_{\rm e}c^2\dot{N}_{\rm p}$, 
and $L_{\rm sec, B\gamma}\equiv2\gamma_{\rm s,syn}m_{\rm e}c^2\dot{N}_{\rm p}N_{\gamma}E_{\rm cur}/E_{\rm esc}$, 
respectively. 
See text details.}
   \label{image}
  \end{center}
 \end{figure*}

We consider synchrotron radiation emitted from created pairs via $\gamma\gamma$ or B$\gamma$ 
pair creation process in the magnetosphere. 
In this paper, we use the term `primary particles' as electrons or positrons that are accelerated 
in the magnetosphere and emit curvature photons whose energy is high enough to convert pairs. 
We also use `secondary particles'
as electrons and positrons created outside of the acceleration region and emit synchrotron radiation. 
Since we are interested in particles which emit the synchrotron radiation, 
`secondary particles' include second and higher generation particles which could emit synchrotron radiation. 

In our model, we make three main assumptions that 
(a) the physical quantities at the regions of the particle acceleration, creation and emission are described 
    as a function of the radius $r$ from the centre of the neutron star to the radius of the light cylinder $R_{\rm lc}$, 
(b) the energy source is only the rotational energy of a neutron star and 
other sources such as the magnetic field dissipation do not contribute to the particle energy flux, and 
(c) the dipole component of the magnetic field dominates near the light cylinder, 
where the radius of the light cylinder is
\begin{eqnarray}\label{R_lc}
R_{\rm lc}=\frac{Pc}{2\pi},
\end{eqnarray} 
where $c$ is the speed of light.
Note that although we only consider the region inside the light cylinder as in assumption (a), 
the model with $r\sim R_{\rm lc}$ is applicable to the the case where significant dissipation of 
Poynting flux takes place close behind the light cylinder ($\sim1-2R_{\rm lc}$) as indicated 
by recent global models \citep{B15, CPPS15}. 
Moreover, the synchrotron luminosity becomes maximum at $R=R_{\rm lc}$
if our model extends outside the light cylinder as discussed in section \ref{sec:discussion}. 
From assumption (b), the total energy flux of primary particles is limited by the spin-down luminosity. 
We introduce the dimensionless conversion efficiency $\eta(\le1)$ from the spin-down luminosity $L_{\rm sd}$ 
to the energy flux of the primary particles as a model parameter (see also equation \ref{dotN_p}).

The radius, mass and dipole magnetic moment of a neutron star are $R_{\rm ns}=10^6$ cm, $M_{\rm ns}=1.4M_{\odot}$, 
and $\mu_{\rm mag}^2\equiv(B_{\rm s}R_{\rm ns}^3/2)^2=3Ic^3P\dot{P}/8\pi^2$, respectively, 
where $I=(2/5)M_{\rm ns}R_{\rm ns}^2$ is moment of inertia. 
For the $\gamma\gamma$ pair creation process, 
we consider the thermal X-ray photons from the heated polar cap 
as the seed photons to create the secondary pairs, 
while we neglect the thermal photons from the entire surface and the synchrotron photons.
This is reasonable approximation except for very young pulsar such as the Crab pulsar \citep[e.g., ][]{YP04}. 
The synchrotron luminosity in ingoing case, where synchrotron emitting pairs are created 
by the curvature photons emitted by the ingoing primary particles,
is lower than the luminosity in outgoing case as shown in previous studies \citep{KT14, KT15}.
We only consider the synchrotron radiation from pairs created 
by the curvature photons emitted by the outgoing primary particles. 

We consider two pair creation processes, $\gamma\gamma$ and B$\gamma$ pair creations, separately. 
The schematic pictures of two scenarios are shown in figure \ref{image}.
In $\gamma\gamma$ pair creation scenario (upper panel), 
the curvature photons (blue arrow) collide with the thermal photons (green arrow) from the heated polar cap. 
Then, the secondary electrons and positrons are created and emit the synchrotron radiation (red arrow). 
The particle acceleration and $\gamma\gamma$ pair creation regions reside near the light cylinder 
as is considered in the outer gap model \citep[e.g., ][]{CHR86}. 
The thermal emission from the heated polar cap are detected in soft X-ray observations 
\citep[e.g., ][]{HR93, ZP04}, 
so that we can easily estimate the optical depth for $\gamma\gamma$ pair creation. 
Since the optical depth for the propagation of curvature photons 
is typically much lower than unity (see section \ref{sec:luminosity})
at the outer magnetosphere, most curvature photons escape from the magnetosphere. 
These pulsars are detected as $\gamma$-ray pulsars \citep[e.g., ][]{Ab13}. 
In $\gamma\gamma$ scenario, 
we use the typical observed energy of $\gamma$-ray pulsars $E_{\rm cur}$ as a parameter, 
to obtain the Lorentz factor of the primary particles as, 
\begin{eqnarray}\label{gamma_p}
\gamma_{\rm p,E}\equiv\gamma_{\rm p}(E_{\rm cur})=\left(\frac{4\pi}{0.87}\frac{E_{\rm cur}}{h}\frac{R_{\rm cur}}{c}\right)^{1/3},
\end{eqnarray}
where $R_{\rm cur}$ is the curvature radius, and $h$ is the Planck constant.
The Lorentz factor $\gamma_{\rm p,E}$ should be lower than the maximum Lorentz factor 
determined by the full potential drop across the polar cap surface,
\begin{eqnarray}\label{gamma_p,max}
\gamma_{\rm p,max}=\frac{2\pi^2eB_{\rm s}R_{\rm ns}^3}{m_{\rm e}c^4P^2},
\end{eqnarray}
where $m_{\rm e}$ and $e$ is the mass and the charge of an electron \citep{GJ69}.
Thus, we use $\gamma_{\rm p}=\min\{\gamma_{\rm p,E}, \gamma_{\rm p,max}\}$ 
as the Lorentz factor of the primary particles. 
Note that even if we consider the non-dipole components such as toroidal and higher order poloidal components 
near the surface as considered in some authors \citep[e.g., ][]{RS75, GV14, SMG15}, 
the potential drop across the polar cap is determined by the dipole magnetic field 
as long as assumption (c) is satisfied. 
Only low-$L_{\rm sd}$ pulsars ($L_{\rm sd}\lesssim10^{34}$ erg s$^{-1}$) satisfy the condition 
$\gamma_{\rm p,max}<\gamma_{\rm p,E}$ with $E_{\rm cur}\sim$ GeV. 
The energy of primary particles could reach almost whole potential drop 
for such low-$L_{\rm sd}$ pulsars \citep{TC09}.

In B$\gamma$ pair creation scenario (lower panel in figure \ref{image}), 
the pair creation occurs near the stellar surface via the interaction 
between the photons and the strong magnetic field.
The curvature (blue arrow) and high-energy synchrotron photons (red arrows) convert to
pairs in B$\gamma$ process.  
Since most of curvature photons emitted from the primary particles 
are efficiently absorbed via B$\gamma$ pair creation process \citep[e.g., ][]{DH82}, 
equation (\ref{gamma_p}) is not available to estimate the Lorentz factor of the primary particles $\gamma_{\rm p}$.
We use the maximum Lorentz factor $\gamma_{\rm p,max}$ in equation (\ref{gamma_p,max}) 
as the Lorentz factor of the primary particles, 
although the synchrotron luminosity does not depend on the Lorentz factor of 
primary particles for most pulsars in our model as discussed in section \ref{sec:results}. 
For the cascade at the inner region, 
we assume that all photons with energy higher than the energy $E_{\rm esc}$ converts to pairs, 
where $E_{\rm esc}$ is the maximum escapable photon energy 
and is described as a function of $r$ (see equation \ref{E_esc}). 
The maximum number of photons which convert to pairs is about $E_{\rm cur}/E_{\rm esc}$ times larger than 
the number of curvature photons with energy $E_{\rm cur}$ from the primary particles. 
This is an ideal cascade case considered in \citet{KT14, TH15}. 
We use this number of photons to estimate the number of the synchrotron emitting particles 
(see equation \ref{dotN_s}). 

In B$\gamma$ scenario, 
we consider two cases for the magnetic field near the stellar surface.
First, only the dipole component dominates at the entire magnetosphere.  
Second, the non-dipole components significantly contribute to the magnetic field at the emission region.
In the latter case, 
we use the simple description to the magnetic field at the emission region as,
\begin{eqnarray}\label{zeta_B}
B\equiv\zeta_{\rm B}B_{\rm s}(r/R_{\rm ns})^{-3},
\end{eqnarray} 
where $\zeta_{\rm B}(\ge1)$ is a model parameter.
Equation (\ref{zeta_B}) reduces to dipole field in the limit $\zeta_{\rm B}=1$.
The configuration of the magnetic field has also effects on 
the curvature radius of the field line $R_{\rm cur}$ and the pitch angle of secondary particles $\alpha$
in our model. 
We use the following approximated formula \citep[e.g., ][]{Tang08, KT14},
\begin{eqnarray}\label{R_cur}
R_{\rm cur}\sim\left\{ \begin{array}{ll}
\sqrt{rR_{\rm lc}} & ~~~({\rm dipole}) \\
 & \\
r & ~~~({\rm non-dipole}), \\
\end{array} \right. 
\end{eqnarray}
and
\begin{eqnarray}\label{alpha}
\alpha\sim\left\{ \begin{array}{ll}
\sqrt{r/R_{\rm lc}} & ~~~({\rm dipole}) \\
 & \\
\alpha_0 & ~~~({\rm non-dipole}), \\
\end{array} \right.
\end{eqnarray}
where the pitch angle $\alpha_0(\le1)$ is a model parameter.

\subsection{Analytical Description of Synchrotron Luminosity}
\label{sec:luminosity}

In our model, 
the newly created secondary particles in the magnetosphere have non-zero pitch angle $\alpha$ (equation \ref{alpha}).
The secondary particles emit synchrotron radiation. 
Taking into account the radiation cooling effect, 
the effective number of the synchrotron emitting particles at the emission region $r$ is 
$\sim\dot{N}_{\rm s}\times\min\{t_{\rm ad}, t_{\rm cool, syn}\}$, 
where $\dot{N}_{\rm s}$ is the total number flux of the secondary particles, 
and two timescales $t_{\rm ad}$ and $t_{\rm cool,syn}$ are the advection and the synchrotron cooling timescales, respectively.
The advection timescale at the emission region $r$ is given by
\begin{eqnarray}\label{t_ad}
t_{\rm ad}\sim\frac{r}{c},
\end{eqnarray}
and the cooling timescale of synchrotron radiation with the characteristic energy $h\nu_{\rm obs}$ is given by
\begin{eqnarray}\label{t_cool,syn}
t_{\rm cool, syn}\sim\frac{\gamma_{\rm s,syn}\alpha m_{\rm e}c^2}{P_{\rm syn}}, 
\end{eqnarray}
where $P_{\rm syn}$ is the power of the synchrotron radiation for each particle
\footnote{This classical formula for synchrotron power is not valid for $\gamma\alpha B/B_{\rm q}>0.1$ 
\citep[e.g., ][]{HL06}. 
However, we consider the characteristic energy of synchrotron radiation, which is much lower than 
the electron rest mass energy so that condition $\gamma\alpha B/B_{\rm q}\ll0.1$ is satisfied.},
\begin{eqnarray}\label{P_syn}
P_{\rm syn}=\frac{2e^4B^2\alpha^2}{3c^3m_{\rm e}^2}\gamma_{\rm s,syn}^2,  
\end{eqnarray}
and $\gamma_{\rm s,syn}$ is the Lorentz factor of the secondary particles 
which emit the synchrotron radiation with the characteristic energy $h\nu_{\rm obs}$, 
\begin{eqnarray}\label{gamma_s,syn}
\gamma_{\rm s,syn}=\sqrt{\frac{4\pi}{0.87}\nu_{\rm obs}\frac{m_{\rm e}c}{eB\alpha}}. 
\end{eqnarray}
Using the effective number of secondary particles and the synchrotron power for each particle,  
the luminosity of the synchrotron radiation from the secondary pairs is described as,
\begin{eqnarray}\label{L_syn}
L_{\rm syn}\sim P_{\rm syn}\dot{N}_{\rm s}\min\{t_{\rm ad}, t_{\rm cool, syn}\}.
\end{eqnarray}

We model the radiative transfer for the curvature photons analytically to calculate 
the number flux of the secondary pairs, $\dot{N}_{\rm s}$. 
The number flux of the secondary pairs in $\gamma\gamma$ and B$\gamma$ scenarios is described by 
\begin{eqnarray}\label{dotN_s}
\dot{N}_{\rm s}\sim2\dot{N}_{\rm p}N_{\gamma}\times\left\{ \begin{array}{ll}
\min\{\tau_{\gamma\gamma},1\} & ~~~~ (\gamma\gamma), \\
 & \\
\frac{E_{\rm cur}}{E_{\rm esc}} & ~~~~ ({\rm B}\gamma), \\
\end{array} \right. 
\end{eqnarray}
where $\dot{N}_{\rm p}$ is the total number fux of the primary particles, 
$N_{\gamma}$ is the number of curvature photons emitted from one primary particle, and 
$\tau_{\gamma\gamma}$ is the optical depth for $\gamma\gamma$ pair creation. 
The factor 2 on the right-hand side of equation (\ref{dotN_s}) accounts for electron and positron. 
The optical depth for the curvature photons in $\gamma\gamma$ pair creation process is given by
\begin{eqnarray}\label{tau_gammagamma}
\tau_{\gamma\gamma}\sim\frac{L_{\rm pc}}{4\pi r^2cE_{\rm pc}}\sigma_{\gamma\gamma}(1-\cos\theta_{\rm col}) r, 
\end{eqnarray}
where $L_{\rm pc}$ is the luminosity of the thermal emission from the heated polar cap, 
$E_{\rm pc}=2.82kT_{\rm pc}$ and $T_{\rm pc}$ are the energy and temperature of the thermal photons, 
$k$ is Boltzmann constant, $\sigma_{\gamma\gamma}\sim0.2\sigma_{\rm T}$ and $\sigma_{\rm T}$ 
are the cross sections for the $\gamma\gamma$ pair creation and the Thomson scattering, and 
$\theta_{\rm col}$ is the collision angle of the curvature and thermal photons, which approximately described by
\begin{eqnarray}\label{theta_col}
1-\cos\theta_{\rm col}\sim\frac{1}{2}\left(\frac{r}{R_{\rm cur}}\right)^2\sim\frac{1}{2}\left(\frac{r}{R_{\rm lc}}\right).
\end{eqnarray}
The soft X-ray observations suggest that the thermal luminosity from the heated polar cap $L_{\rm pc}$ is 
approximately proportional to the spin-down luminosity, $L_{\rm pc}\sim10^{-3}L_{\rm sd}$ \citep{BT97, B09}.
We use the normalized luminosity $\epsilon_{\rm pc}\equiv L_{\rm pc}/L_{\rm sd}$ instead of $L_{\rm pc}$.
For typical $\gamma$-ray pulsars, the optical depth is 
$\tau_{\gamma\gamma}\sim2.7\times10^{-4}(\epsilon_{\rm pc}/10^{-3})(T_{\rm pc}/10^6{\rm K})^{-1}(P/0.1{\rm s})^{-1}$, 
much smaller than unity. 
Only a part of the curvature photons converts to the pairs. 
On the other hand, in B$\gamma$ pair creation, because of the high efficiency of the pair conversion, 
we assume that all photons with energy higher than $E_{\rm esc}$ convert to pairs in the electromagnetic cascade 
\citep{KT14, TH15}. 
The maximum energy of the escapable photon is described by \citep[e.g., ][]{E66}
\begin{eqnarray}\label{E_esc}
E_{\rm esc}=2m_{\rm e}c^2\chi_{\min}\frac{B_{\rm q}}{B_{\perp}}, 
\end{eqnarray}
where the magnetic fields $B_{\rm q}=m_{\rm e}^2c^3/e\hbar\sim4.4\times10^{13}{\rm G}$ 
and $B_{\perp}=B\sin\theta_{\rm B\gamma}$, and we take the critical value as $\chi_{\min}=1/15$ following \citet{RS75}
\footnote{Equation (\ref{E_esc}) is not valid for $B>0.1B_{\rm q}$ \citep[e.g., ][]{DH83}. 
However, from inequality (\ref{frequency_condition}) discussed in section \ref{sec:constraints}, 
the magnetic field is limited $B\lesssim0.1\alpha B_{\rm q}$ at the emission region 
as long as we consider the energy range $h\nu_{\rm obs}\lesssim50$ keV.}
.
The angle $\theta_{\rm B\gamma}$, which is the angle between the direction of the propagation 
for a curvature photon and the magnetic field, is comparable to the pitch angle of secondary pairs (equation \ref{alpha}). 
Then, we use the approximation $\theta_{\rm B\gamma}\sim\alpha$. 

We model the primary particles that emit the curvature radiation to estimate the number flux $\dot{N}_{\rm p}$ and 
the number of curvature photons $N_{\gamma}$. 
For the total number flux of the primary particles $\dot{N}_{\rm p}$ with the Lorentz factor $\gamma_{\rm p}$, 
we introduce a parameter $\eta$ which corresponds to the conversion efficiency 
from the spin-down luminosity $L_{\rm sd}$ as 
\begin{eqnarray}\label{dotN_p}
\dot{N}_{\rm p}=\frac{\eta L_{\rm sd}}{\gamma_{\rm p}m_{\rm e}c^2}. 
\end{eqnarray}
From assumption (b) in section \ref{sec:assumptions}, the efficiency must be smaller than unity, $\eta\le1$. 
In order to estimate the number of curvature photons for a particle $N_{\gamma}$, 
we consider two timescales, the advection $t_{\rm ad}$ (equation \ref{t_ad}) 
and the curvature cooling timescales $t_{\rm cool, cur}$.
The curvature cooling timescale is given by
\begin{eqnarray}\label{t_cool,cur}
t_{\rm cool, cur}\sim\frac{\gamma_{\rm p}m_{\rm e}c^2}{P_{\rm cur}}, 
\end{eqnarray}
where $P_{\rm cur}$ is the power of curvature radiation from a particle, 
\begin{eqnarray}\label{P_cur}
P_{\rm cur}=\frac{2e^2c}{3R_{\rm cur}^2}\gamma_{\rm p}^4.
\end{eqnarray}
The number of the curvature photons for a particle $N_{\gamma}$ is estimated by
\begin{eqnarray}\label{N_gamma}
N_{\gamma}\sim\frac{P_{\rm cur}}{E_{\rm cur}}\min\{t_{\rm cool,cur},t_{\rm ad}\}.
\end{eqnarray}
Substituting equations (\ref{P_syn}), (\ref{gamma_s,syn}), (\ref{dotN_s} - \ref{N_gamma}) into equation (\ref{L_syn}), 
we obtain the luminosity of the synchrotron radiation as a function of $r$. 
For the derived values, model parameters is only $\eta$ in $\gamma\gamma$ and B$\gamma$ scenario 
in the dipole dominant case. 
If we take into account the effects of the non-dipole field for the strength of the local magnetic field 
(equation \ref{zeta_B}), 
the pitch angle (equation \ref{alpha}), and the curvature radius (equation \ref{R_cur}) in B$\gamma$ scenario, 
the derived synchrotron luminosity also depends on the additional model parameters, $\alpha_0$ and $\zeta_{\rm B}$. 
The values of $\nu_{\rm obs}, E_{\rm cur}, \epsilon_{\rm pc}, T_{\rm pc}, B_{\rm s}$ and $L_{\rm sd}$ 
are taken from the observations.

\subsection{Constraints on Emission Region}
\label{sec:constraints}

There are some constraints on the emission region $r$ in our synchrotron radiation model.
We assume that the emission region resides in the magnetosphere, 
\begin{eqnarray}\label{magnetosphere}
R_{\rm ns}<r<R_{\rm lc}.
\end{eqnarray}
In addition, we consider following four conditions in order to constrain the emission region $r$.

First condition is that the Lorentz factor of the created secondary particles 
from the curvature photons $\gamma_{\rm s,pair}$ has to be larger than 
the Lorentz factor of particles $\gamma_{\rm s,syn}$ which emit the synchrotron radiation 
with the characteristic energy $h\nu_{\rm obs}$, 
\begin{eqnarray}\label{energy_condition}
\gamma_{\rm s,pair}>\gamma_{\rm s,syn},  
\end{eqnarray}
where $\gamma_{\rm s,pair}$ is the Lorentz factor of the secondary pairs from a photon energy $E_{\rm cur}$,
\begin{eqnarray}\label{gamma_s,pair}
\gamma_{\rm s,pair}=\frac{E_{\rm cur}}{2m_{\rm e}c^2}.
\end{eqnarray}
Since the Lorentz factor for a given frequency $\nu_{\rm obs}$ is $\gamma_{\rm s,syn}\propto r^{3/2}\alpha^{-1/2}$ 
(equation \ref{gamma_s,syn}), energy condition (\ref{energy_condition}) 
gives the upper limit on the emission region, $r\le r_{\gamma{\rm syn}}$.

Second condition is the validity of the synchrotron approximation, $\gamma_{\rm s,syn}\alpha>1$. 
The condition indicates that the observed frequency $\nu_{\rm obs}$ 
should be higher than 
the low-energy turnover frequency for the synchrotron radiation \citep{OS70, RD99},
\begin{eqnarray}\label{frequency_condition}
\nu_{\rm obs}>\frac{eB}{2\pi m_{\rm e}c\alpha}.
\end{eqnarray}
Condition (\ref{frequency_condition}) gives the lower limit on the emission region, $r\ge r_{\rm ct}$.

Third and fourth conditions come from the pair creation thresholds. 
The $\gamma\gamma$ pair creation threshold is described by
\begin{eqnarray}\label{gamma-gamma_condition}
(1-\cos\theta_{\rm col})E_{\rm pc}E_{\rm cur}>2(m_{\rm e}c^2)^2.
\end{eqnarray}
Under the dipole magnetic field configuration, 
the collision angle $\theta_{\rm col}$ becomes large toward the large distance 
from the neutron star (equation \ref{theta_col}).
Then, condition (\ref{gamma-gamma_condition}) gives the lower limit, $r\ge r_{\gamma\gamma}$.
The B$\gamma$ pair creation threshold is described by \citep{E66} 
\begin{eqnarray}\label{B-gamma_condition}
\frac{E_{\rm cur}}{2m_{\rm e}c^2}\frac{B_{\perp}}{B_{\rm q}}>\chi_{\min}. 
\end{eqnarray}
Since the occurrence of the B$\gamma$ pair production requires the strong magnetic field,
condition (\ref{B-gamma_condition}) gives upper limit $r\le r_{\rm B\gamma}$.

From conditions (\ref{magnetosphere}-\ref{gamma-gamma_condition}), 
the range of the emission region in $\gamma\gamma$ pair creation scenario is 
\begin{eqnarray}\label{gamma-gamma_range}
\max\{R_{\rm ns}, r_{\rm ct}, r_{\gamma\gamma}\}<r<\min\{R_{\rm lc},r_{\gamma{\rm syn}}\}.
\end{eqnarray}
Within typical parameter ranges, the relation $\max\{R_{\rm ns}, r_{\rm ct}, r_{\gamma\gamma}\}=r_{\gamma\gamma}$ would be satisfied.
Then, the allowed range of the emission region is 
$r_{\gamma\gamma}<r<\min\{R_{\rm lc},r_{\gamma{\rm syn}}\}$
for most pulsars.

In B$\gamma$ pair creation scenario, the range of the emission region is 
\begin{eqnarray}\label{B-gamma_range}
\max\{R_{\rm ns}, r_{\rm ct}\}<r<\min\{R_{\rm lc},r_{\gamma{\rm syn}},r_{\rm B\gamma}\} 
\end{eqnarray}
from conditions (\ref{magnetosphere}-\ref{frequency_condition}) and (\ref{B-gamma_condition}). 
Within the typical ranges of the parameters, the relations 
$\max\{R_{\rm ns}, r_{\rm ct}\}=r_{\rm ct}$, $r_{\gamma{\rm syn}}>R_{\rm lc}$, and
$r_{\gamma{\rm syn}}>r_{\rm B\gamma}$ would be satisfied.
Then, the range of the emission region is $r_{\rm ct}<r<\min\{R_{\rm lc},r_{\rm B\gamma}\}$.

\section{RESULTS}
\label{sec:results}

In this section, we calculate the luminosity of the synchrotron radiation based on the model described in previous section.
In section \ref{sec:efficiency}, we derive the efficiency of the synchrotron radiation relative to the spin-down luminosity 
as a function of $r$ from equation (\ref{L_syn}). 
In section \ref{sec:deathline}, we provide the allowed range of the emission region $r$ and the death lines 
for each scenario from the constraints described in section \ref{sec:constraints}.
Using the efficiency of the synchrotron radiation as a function of $r$ and the constraints on $r$, 
we provide the maximum luminosity in the allowed emission region and 
its dependence on the spin-down luminosity in section \ref{sec:maximum}.
Comparing with the observed non-thermal luminosity, 
we provide the required range of the efficiency parameter $\eta$.
In section \ref{sec:ratio}, we derive the ratio of the curvature to the synchrotron luminosities, 
which do not depend on the distance from the observer and the model parameter $\eta$. 
Then, we compare with the flux ratios of observed $\gamma$-ray to X-ray and optical.
Hereafter, we use $Q_x\equiv Q/10^x$ in cgs units, $E_{\rm cur, GeV}\equiv E_{\rm cur}/1$ GeV, 
and $h\nu_{\rm keV}\equiv h\nu/1$ keV. 

\subsection{Synchrotron Efficiency}
\label{sec:efficiency}

Before the detailed calculations, we see the main control parameters of the efficiency of the synchrotron radiation 
$\epsilon_{\rm syn}\equiv L_{\rm syn}/L_{\rm sd}$ in the conditions where the radiative cooling timescales 
are smaller than the advection timescale in equations (\ref{dotN_s}) and (\ref{N_gamma}) 
($t_{\rm cool,syn}<t_{\rm ad}$ and $t_{\rm cool,cur}<t_{\rm ad}$).

In $\gamma\gamma$ pair creation scenario, 
the efficiency $\epsilon_{\rm syn}$ is derived from equation (\ref{L_syn}) as, 
\begin{eqnarray}\label{epsilon_syn,gamma-gamma}
\epsilon_{\rm syn}\sim\eta\tau_{\gamma\gamma}\frac{\gamma_{\rm s,syn}\alpha}{\gamma_{\rm s,pair}}.
\end{eqnarray}
Near the light cylinder, $r\sim R_{\rm lc}$, the pitch angle $\alpha$ becomes order unity (equation \ref{alpha}).
For the typical $\gamma$-ray pulsars, the optical depth is $\tau_{\gamma\gamma}\sim10^{-4}-10^{-2}$ 
(equation \ref{tau_gammagamma})
and the energy ratio $\gamma_{\rm s,syn}/\gamma_{\rm s,pair}\sim(0.1-1)\nu_{\rm obs,keV}^{1/2}$ 
(equations \ref{gamma_s,syn} and \ref{gamma_s,pair}). 
Then, the efficiency of the synchrotron radiation is roughly 
$\epsilon_{\rm syn}\sim10^{-5}-10^{-2}\eta\nu_{\rm obs,keV}^{1/2}$ 
which is roughly consistent with observed non-thermal X-ray efficiency \citep[e.g., ][]{Kar12} 
if the conversion efficiency is an order of unity, $\eta\sim1$.

In B$\gamma$ pair creation scenario, 
the efficiency is
\begin{eqnarray}\label{epsilon_syn,B-gamma}
\epsilon_{\rm syn}\sim\eta\frac{\gamma_{\rm s,syn}\alpha m_{\rm e}c^2}{E_{\rm esc}}.
\end{eqnarray}
This efficiency does not depend on the Lorentz factor of primary particles $\gamma_{\rm p}$, 
unless $\gamma_{\rm p}$ is too low to create the pairs which emit synchrotron radiation.
Near the stellar surface $r\sim R_{\rm ns}$, 
since the energy ratio is 
$\gamma_{\rm s,syn}m_{\rm e}c^2/E_{\rm esc}\sim10^{-2}\alpha^{1/2}\zeta_{\rm B}^{1/2}\nu_{\rm obs,keV}^{1/2}$, 
the efficiency is $\epsilon_{\rm syn}\sim10^{-2}\eta\alpha^{3/2}\zeta_{\rm B}^{1/2}\nu_{\rm obs,keV}^{1/2}$.
If we consider the high conversion efficiency $\eta\sim1$ 
and the non-dipole magnetic field with the pitch angle $\alpha\sim1$, 
the synchrotron efficiency could be $\epsilon_{\rm syn}\sim10^{-2}\zeta_{\rm B}^{1/2}\nu_{\rm obs,keV}^{1/2}$. 

The conditions, $t_{\rm cool,syn}<t_{\rm ad}$ and $t_{\rm cool,cur}<t_{\rm ad}$, are not always satisfied. 
Then, the synchrotron efficiency becomes reduced. 
In what follows, we calculate the efficiency of the synchrotron radiation comparing timescales 
$t_{\rm ad}, t_{\rm cool,syn},$ and $t_{\rm cool,cur}$. 
First, we compare the synchrotron cooling and advection timescales, $t_{\rm cool, syn}$ and $t_{\rm ad}$, 
to calculate the synchrotron luminosity $L_{\rm syn}$ from equation (\ref{L_syn}).
The normalized radius $r_{\rm eq,syn,6}\equiv r_{\rm eq,syn}/10^6{\rm cm}$ where the synchrotron cooling timescale 
equals to the advection timescale, $t_{\rm cool, syn}=t_{\rm ad}$, is 
\begin{eqnarray}\label{r_eq,syn}
r_{\rm eq,syn,6}\sim\left\{ \begin{array}{ll}
8.6\times10^2\nu_{\rm obs, keV}^{2/13}B_{\rm s, 12}^{11/26}L_{\rm sd,31}^{1/52} & \\ 
~~~~~~~~~~~~~~~~~~~~~~~~ ({\rm dipole}), & \\
 & \\
9.7\times10^2\alpha_0^{1/7}\zeta_{\rm B}^{3/7}\nu_{\rm obs,keV}^{1/7}B_{\rm s,12}^{3/7} & \\
~~~~~~~~~~~~~~~~~~~~~~~~ ({\rm non-dipole}). & \\
\end{array} \right. 
\end{eqnarray} 
At the inner region $r<r_{\rm eq,syn}$, 
we should consider the synchrotron cooling timescale ($t_{\rm cool,syn}<t_{\rm ad}$)
to calculate the luminosity $L_{\rm syn}$ from equation (\ref{L_syn}). 
In B$\gamma$ scenario, the emission region resides near the stellar surface 
(section \ref{sec:assumptions}, figure \ref{image}), so that 
condition $r<r_{\rm eq,syn}$ ($t_{\rm cool, syn}<t_{\rm ad}$) is satisfied. 
We only consider $\min\{t_{\rm cool,syn}, t_{\rm ad}\}=t_{\rm cool,syn}$ in B$\gamma$ scenario.
On the other hand, in $\gamma\gamma$ scenario, 
the emission region could reside at the outer region of the magnetosphere 
(section \ref{sec:assumptions}, figure \ref{image}), 
so that the emission region could reside $r>r_{\rm eq,syn}$ ($t_{\rm cool,syn}>t_{\rm ad}$)
for some pulsars with the low spin-down luminosity. 
Note that the radius $r_{\rm eq,syn}$ in equation (\ref{r_eq,syn}) does not depend on 
whether equation (\ref{gamma_p}) or (\ref{gamma_p,max}) is used as the Lorentz factor of the primary particles.

Next, we compare the curvature cooling timescale $t_{\rm cool, cur}$ and advection timescale $t_{\rm ad}$  
to calculate the number of $\gamma$-ray photons $N_{\gamma}$ from equation (\ref{N_gamma}).
Using the dipole magnetic field and the Lorentz factor of the primary particles 
$\gamma_{\rm p,E}$ (equation \ref{gamma_p}), 
the condition $t_{\rm cool, cur}<t_{\rm ad}$ gives the lower limit $r>r_{\rm eq,cur}$ on the radius, 
\begin{eqnarray}\label{r_eq,cur-E_cur}
r_{\rm eq,cur,6}\sim
9.9E_{\rm cur,GeV}^{-2}B_{\rm s,12}^{1/2}L_{\rm sd,31}^{-1/4} \nonumber \\
~~~~~~~~~~~~~~ (\gamma_{\rm p,E}, {\rm dipole}).
\end{eqnarray}
In $\gamma\gamma$ scenario, the emission region resides near the light cylinder, $r\sim R_{\rm lc}$,
so that $r>r_{\rm eq,cur}$ ($t_{\rm cool, cur}<t_{\rm ad}$) is satisfied for most pulsars. 
If we use the Lorentz factor $\gamma_{\rm p,max}$ (equation \ref{gamma_p,max}), 
condition $t_{\rm cool, cur}<t_{\rm ad}$ does not depend on the radius $r$. 
Condition $t_{\rm cool, cur}<t_{\rm ad}$ constrains on the spin-down luminosity as,
\begin{eqnarray}\label{L_eq,cur}
L_{\rm sd,31}\gtrsim4.0B_{\rm s,12}^{2/7}~~~~~(\gamma_{\rm p,max}, {\rm dipole}),
\end{eqnarray}
which is satisfied for most pulsars. 
This condition is applicable to both $\gamma\gamma$ and B$\gamma$ scenarios.
On the other hand, if we consider the non-dipole components of the magnetic field at the emission region 
and use the Lorentz factor of the primary particles $\gamma_{\rm p,max}$ as considered in B$\gamma$ scenario, 
condition $t_{\rm cool, cur}<t_{\rm ad}$ gives the upper limit ($r<r_{\rm eq,cur}$) on the radius,
\begin{eqnarray}\label{r_eq,cur-gamma_p,max}
r_{\rm eq,cur,6}\sim4.3\times10^2L_{\rm sd,31}^{3/2} \nonumber \\
~~~~~~~~~~~~~~ (\gamma_{\rm p,max}, {\rm non-dipole}).
\end{eqnarray}
Radius $r_{\rm eq,cur}$ is much larger than the stellar radius $R_{\rm ns}$. 
In B$\gamma$ scenario, the emission region resides near the stellar surface, $r\sim R_{\rm ns}$.
Then, condition $r<r_{\rm eq,cur}$ ($t_{\rm cool,cur}<t_{\rm ad}$) at the emission region should be satisfied.
From conditions (\ref{r_eq,cur-E_cur}-\ref{r_eq,cur-gamma_p,max}), 
we use $\min\{t_{\rm cool,cur}, t_{\rm ad}\}=t_{\rm cool,cur}$ for most pulsars. 

 \begin{figure*}
  \begin{center}
   \includegraphics[width=140mm]{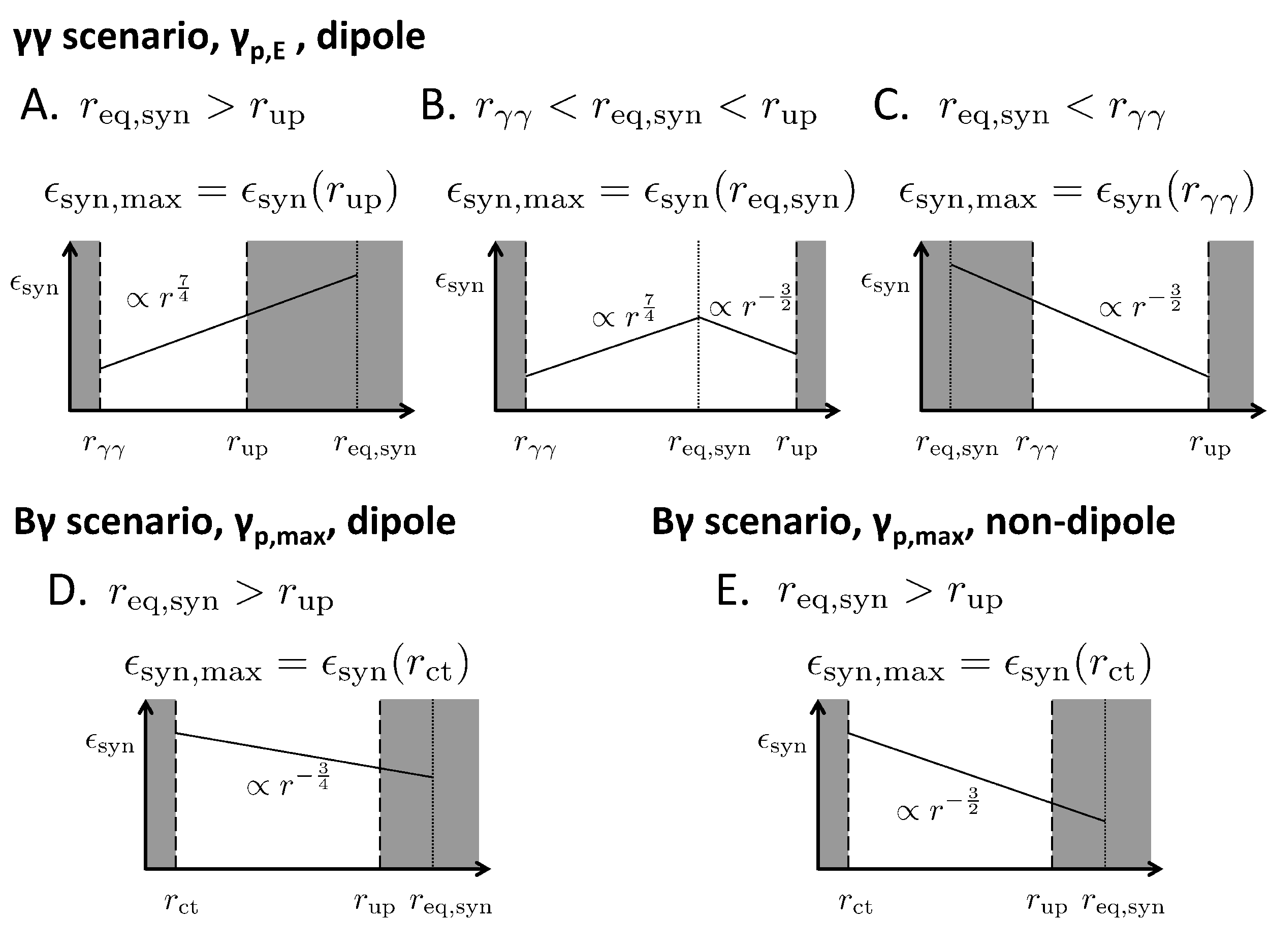}
   \caption{Schematic pictures for the $r$ dependence of the efficiency of 
the synchrotron radiation $\epsilon_{\rm syn}$. 
The upper limits on the emission region are $r_{\rm up}\equiv\min\{R_{\rm lc}, r_{\rm \gamma,syn}\}$ 
for $\gamma\gamma$ scenario (panels A-C), and $r_{\rm up}\equiv\min\{R_{\rm lc}, r_{\rm B\gamma}\}$ 
for B$\gamma$ scenario (panels D and E).}
   \label{schematics}
  \end{center}
 \end{figure*}

Substituting equations (\ref{alpha} - \ref{gamma_s,syn}) and (\ref{dotN_s} - \ref{N_gamma}) 
into equation (\ref{L_syn}), we obtain the efficiencies of synchrotron radiation $\epsilon_{\rm syn}$ 
as a function of the radius $r$.
In $\gamma\gamma$ pair creation scenario, the efficiencies are given by 
\begin{eqnarray}\label{epsilon_syn,gammagamma}
\epsilon_{\rm syn}\sim\left\{ \begin{array}{ll}
1.7\times10^{-14}\eta\nu_{\rm obs, keV}^{1/2}E_{\rm cur, GeV}^{-1} & \\ 
~~~~\times\epsilon_{\rm pc, -3}T_{\rm pc, 6.5}^{-1}B_{\rm s,12}^{-9/8}L_{\rm sd,31}^{21/16}r_6^{7/4} & \\
~~~~~~~~~ (t_{\rm cool, syn}<t_{\rm ad}, \gamma_{\rm p,E}), & \\
 & \\
5.9\times10^{-14}\eta\nu_{\rm obs, keV}^{1/2} & \\ 
~~~~\times\epsilon_{\rm pc, -3}T_{\rm pc, 6.5}^{-1}B_{\rm s,12}^{-7/8}L_{\rm sd,31}^{-5/16}r_6^{9/4} & \\
~~~~~~~~~ (t_{\rm cool, syn}<t_{\rm ad}, \gamma_{\rm p,max}), & \\
 & \\
5.8\times10^{-5}\eta\nu_{\rm obs, keV}E_{\rm cur, GeV}^{-1} & \\ 
~~~~\times\epsilon_{\rm pc, -3}T_{\rm pc,6.5}^{-1}B_{\rm s,12}^{1/4}L_{\rm sd,31}^{11/8}r_6^{\rm -3/2} &  \\
~~~~~~~~~ (t_{\rm cool, syn}>t_{\rm ad}, \gamma_{\rm p,E}), & \\
 & \\
2.1\times10^{-4}\eta\nu_{\rm obs, keV} & \\ 
~~~~\times\epsilon_{\rm pc, -3}T_{\rm pc,6.5}^{-1}B_{\rm s,12}^{1/2}L_{\rm sd,31}^{-1/4}r_6^{\rm -1} &  \\
~~~~~~~~~ (t_{\rm cool, syn}>t_{\rm ad}, \gamma_{\rm p,max}). & \\
\end{array} \right. 
\end{eqnarray}
The dependence of the efficiency $\epsilon_{\rm syn}$ in $\gamma\gamma$ scenario on the radius $r$ 
is shown in the upper panels of figure \ref{schematics}. 
Although we only show the case $\gamma_{\rm p}=\gamma_{\rm p,E}$ in the upper panels of figure \ref{schematics},
the following trends are the same in the case $\gamma_{\rm p}=\gamma_{\rm p,max}$. 
Under condition $t_{\rm cool,syn}<t_{\rm ad}$ $(r<r_{\rm eq,syn})$, 
the efficiency $\epsilon_{\rm syn}$ increases toward the outer region (panels A and B in figure \ref{schematics}). 
Since the pitch angle $\alpha$ and the energy ratio 
$\gamma_{\rm s,syn}/\gamma_{\rm s,pair}$ increase toward the outer region, 
the efficiency $\epsilon_{\rm syn}$ tends to increase with the increase of $r$.
In the outer region $r>r_{\rm eq,syn}$ ($t_{\rm cool,syn}>t_{\rm ad}$), 
the fraction of the total momentum of secondary particles, which loses via synchrotron radiation, 
decreases toward the large distance $r$. 
Thus, the efficiency $\epsilon_{\rm syn}$ decreases toward the large distance $r$ 
(panels B and C in figure \ref{schematics}).

In a similar way as in equation (\ref{epsilon_syn,gammagamma}), 
the efficiencies in B$\gamma$ pair creation scenario are given by
\begin{eqnarray}\label{epsilon_syn,Bgamma}
\epsilon_{\rm syn}\sim\left\{ \begin{array}{ll}
2.7\times10^{-4}\eta\nu_{\rm obs, keV}^{1/2} & \\  
~~~~\times B_{\rm s,12}^{1/8}L_{\rm sd, 31}^{3/16}r_6^{\rm -3/4} & \\
~~~~~~~~~ (t_{\rm cool, syn}<t_{\rm ad}, {\rm dipole}), & \\
 & \\
1.5\times10^{-1}\alpha_0^{3/2}\eta\zeta_{\rm B}^{1/2} & \\ 
~~~~\times\nu_{\rm obs, keV}^{1/2}B_{\rm s,12}^{1/2}r_6^{\rm -3/2} &  \\
~~~~~~~~~ (t_{\rm cool, syn}<t_{\rm ad}, {\rm non-dipole}). & \\
\end{array} \right. 
\end{eqnarray}
Schematic pictures for the efficiency in B$\gamma$ scenario as a function of radius $r$ 
is shown in the lower panels of figure \ref{schematics}.
As already mentioned, we only consider $t_{\rm cool,syn}<t_{\rm ad}$ in B$\gamma$ scenario (equation \ref{r_eq,syn}).
The dependence of the efficiency $\epsilon_{\rm syn}$ on $r$ mainly comes from the photon escaping 
energy $E_{\rm esc}$ (equation \ref{E_esc}).
Since the strong magnetic field makes the pair conversion efficient, 
the energy $E_{\rm esc}$ decreases toward the neutron star surface. 
Then, the efficiency $\epsilon_{\rm syn}$ increases toward the inner region 
(panels D and E in figure \ref{schematics}).

\subsection{Allowed Emission Region and Death Lines}
\label{sec:deathline}

 \begin{figure*}
  \begin{center}
   \vspace{10mm}
   \includegraphics[width=115mm, angle=270]{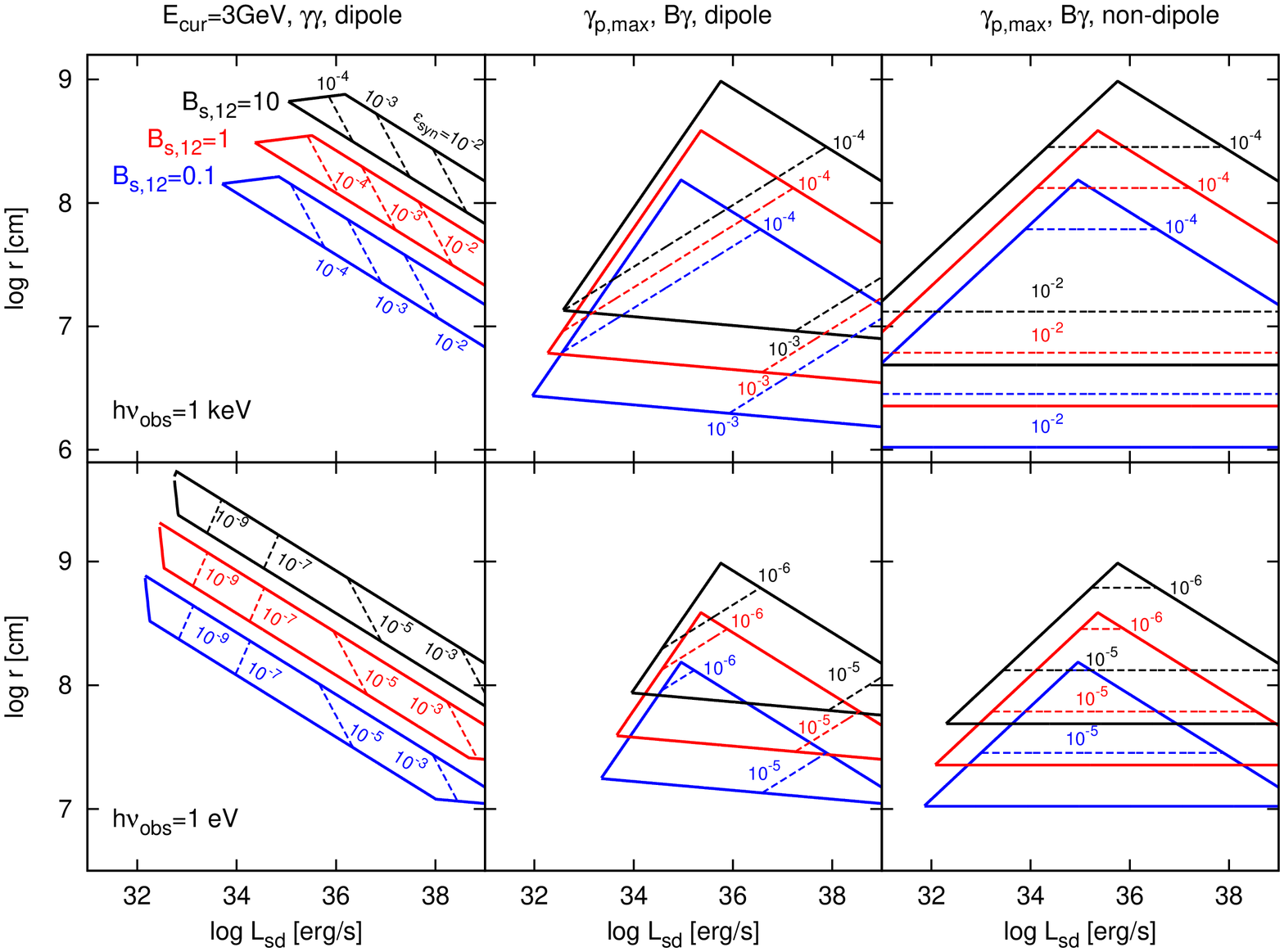}
   \caption{The allowed ranges of the emission region $r$ as functions of the spin-down luminosity 
$L_{\rm sd}$ for the surface dipole field $B_{\rm s}=10^{11}$G (blue), $10^{12}$G (red), and $10^{13}$G (black). 
The allowed ranges are surrounded by solid lines. 
Upper panels show the X-ray emission region ($h\nu_{\rm obs}=1$ keV), 
and lower panels show the optical emission region ($h\nu_{\rm obs}=1$ eV). 
The results in the scenarios of $\gamma\gamma$ with $E_{\rm cur}=3$ GeV, B$\gamma$ with dipole field, 
and B$\gamma$ with non-dipole field are shown in left, middle, and right panels, respectively. 
The dashed lines denote the efficiency of the synchrotron radiation $\epsilon_{\rm syn}$ 
from equations (\ref{epsilon_syn,gammagamma}) and (\ref{epsilon_syn,Bgamma}). 
The parameters are fixed to $\alpha_0=\eta=\zeta_{\rm B}=1$. }
   \label{region}
  \end{center}
 \end{figure*}

In order to derive the maximum luminosity of the synchrotron radiation in the allowed emission region of the magnetosphere, 
we have to consider the constraints on the emission region discussed in section \ref{sec:constraints}.
In this subsection, using the typical values, we derive the limits on the emission region $r$ from the constraints (inequalities \ref{magnetosphere}-\ref{B-gamma_condition}). 
From the energy condition for the secondary particles (\ref{energy_condition}) 
and the B$\gamma$ pair creation threshold (\ref{B-gamma_condition}), 
the upper limits on the emission region $r$ are given by
\begin{eqnarray}\label{r_gammasyn}
r_{\gamma{\rm syn},6}\sim\left\{ \begin{array}{ll}
87\nu_{\rm obs,keV}^{-2/5}E_{\rm cur,GeV}^{4/5}B_{\rm s,12}^{3/10}L_{\rm sd,31}^{1/20}, &  \\ 
~~~~~~~~~~ (\gamma_{\rm p,E}, {\rm dipole}), & \\
 & \\
12\nu_{\rm obs, keV}^{-2/7}B_{\rm s,12}^{1/14}L_{\rm sd,31}^{27/28}, & \\ 
~~~~~~~~~~ (\gamma_{\rm p,max}, {\rm dipole}), & \\
 & \\
71\alpha_0^{1/5}\zeta_{\rm B}^{1/5}\nu_{\rm obs, keV}^{1/5}B_{\rm s,12}^{1/5}L_{\rm sd,31}^{3/5}, & \\
~~~~~~~~~~ (\gamma_{\rm p, max}, {\rm non-dipole}), & \\
\end{array} \right. 
\end{eqnarray}
and
\begin{eqnarray}\label{r_Bgamma}
r_{\rm B\gamma,6}\sim\left\{ \begin{array}{ll}
1.1B_{\rm s,12}^{1/6}L_{\rm sd,31}^{7/12} & \\ 
~~~~~~~~~~ ({\rm dipole}), & \\
  & \\
9.0\alpha_0^{1/4}\zeta_{\rm B}^{1/4}B_{\rm s,12}^{1/4}L_{\rm sd,31}^{3/8} & \\ 
~~~~~~~~~~ ({\rm non-dipole}), & \\
\end{array} \right. 
\end{eqnarray}
respectively. 
On the other hand, 
the lower limits are given by the conditions of the synchrotron approximation (\ref{frequency_condition}) 
and the $\gamma\gamma$ pair creation threshold (\ref{gamma-gamma_condition}), 
\begin{eqnarray}\label{r_ct}
r_{\rm ct,6}\sim\left\{ \begin{array}{ll}
6.7\nu_{\rm obs,keV}^{-2/7}B_{\rm s,12}^{5/14}L_{\rm sd,31}^{-1/28} & \\ 
~~~~~~~~~~ ({\rm dipole}), & \\
  & \\
2.3\alpha_0^{-1/3}\zeta_{\rm B}^{1/3}\nu_{\rm obs,keV}^{-1/3}B_{\rm s,12}^{1/3} & \\
~~~~~~~~~~ ({\rm non-dipole}), & \\
\end{array} \right. 
\end{eqnarray}
and
\begin{eqnarray}\label{r_gammagamma}
r_{\gamma\gamma,6}\sim\left\{ \begin{array}{ll}
    6.4\times10^3T_{\rm pc,6.5}^{-1}E_{\rm cur,GeV}^{-1}B_{\rm s,12}^{1/2}L_{\rm sd,31}^{-1/4} & \\
~~~~~~~~ (\gamma_{\rm p,E}), & \\
 & \\
5.2\times10^8T_{\rm pc,6.5}^{-2}B_{\rm s,12}^{3/2}L_{\rm sd,31}^{-15/4} & \\
~~~~~~~~ (\gamma_{\rm p,max}), & \\
\end{array} \right. 
\end{eqnarray}
respectively. 

In figure \ref{region}, we show the allowed range of the emission regions of the synchrotron radiation 
as the regions surrounded by solid lines as functions of the surface dipole field $B_{\rm s}$ 
and the spin-down luminosity $L_{\rm sd}$. 
We also show the contour lines of the efficiency $\epsilon_{\rm syn}$ with $\eta=1$
as dashed lines in the allowed emission region from equations (\ref{epsilon_syn,gammagamma}) and 
(\ref{epsilon_syn,Bgamma}).

In $\gamma\gamma$ pair creation scenario (left panels in figure \ref{region}), 
we fix the characteristic energy of the curvature photons $E_{\rm cur}=3$ GeV.
The upper limit on the emission region $r$ is determined by the radius of the light cylinder $R_{\rm lc}$ 
(equation \ref{R_lc}) for $h\nu_{\rm obs}=1$ keV (upper panel) and 1 eV (lower panel), 
except for some low spin-down pulsars whose upper limit on the emission region is $r_{\gamma{\rm syn}}$ 
(equation \ref{r_gammasyn}) for $h\nu_{\rm obs}=1$ keV.
The lower limit is determined by the pair creation condition, $r_{\gamma\gamma}$ (equation \ref{r_gammagamma}).
In X-ray band ($h\nu_{\rm obs}=1$ keV)
the efficiency of the synchrotron radiation $\epsilon_{\rm syn}$ becomes the maximum at the light cylinder 
$r=R_{\rm lc}$ (upper left panel in figure \ref{region}). 
For the range of the spin-down luminosity $L_{\rm sd}\lesssim 10^{34}-10^{35}$ erg s$^{-1}$, 
there is no allowed emission region 
in the range $B_{\rm s}\gtrsim10^{11}$ G. 
In optical band ($h\nu_{\rm obs}=1$ eV), 
the radius $r_{\rm eq,syn}$ ($t_{\rm cool, syn}=t_{\rm ad}$; equation \ref{r_eq,syn}) is within the allowed region 
for the pulsars with $L_{\rm sd}\sim10^{34}-10^{36}$ erg s$^{-1}$ (panel B in figure \ref{schematics}).
The narrowness of the allowed emission region for a given $L_{\rm sd}$ 
makes the range of the state in panel B very narrow. 
For pulsars with low spin-down luminosity $L_{\rm sd}\lesssim10^{34}$ erg s$^{-1}$, 
the radius $r_{\rm eq,syn}$ is smaller than the lower limit on the emission region $r_{\gamma\gamma}$ 
(panel C of figure \ref{schematics}).
Then, the maximum efficiency $\epsilon_{\rm syn}$ is given at the region $r=r_{\gamma\gamma}$ 
(lower left panel in figure \ref{region}). 
In fact, the slope of the contours in the lower left panel in figure \ref{region} 
is opposite for $L_{\rm sd}\lesssim10^{34}$ erg s$^{-1}$ and $L_{\rm sd}\gtrsim10^{36}$ erg s$^{-1}$.
Pulsars with $L_{\rm sd}\lesssim10^{32}$ erg s$^{-1}$ cannot emit the synchrotron radiation at optical band
from the secondary particles in our model. 

In B$\gamma$ pair creation scenario (middle and right panels in figure \ref{region}), 
the upper limit on the emission region is determined by the pair creation condition $r_{\rm B\gamma}$ 
for pulsars with $L_{\rm sd}\lesssim10^{35}-10^{36}$ erg s$^{-1}$ or 
the light cylinder $R_{\rm lc}$ for pulsars with $L_{\rm sd}\gtrsim10^{35}-10^{36}$ erg s$^{-1}$.
Since we use the maximum value $\gamma_{\rm p,max}$ as the Lorentz factor of the primary particles, 
the upper limit $r_{\rm B\gamma}$ is large and the allowed region is very broad 
compared with $\gamma\gamma$ scenario.
The lower limit on the emission region is $r_{\rm ct}$ derived 
by the validity condition of the synchrotron approximation (equation \ref{frequency_condition}).
Since the upper limits $r_{\rm B\gamma}$ and $R_{\rm lc}$ do not depend on the observed frequency $\nu_{\rm obs}$, 
the dependence of the area of the allowed region on the frequency $\nu_{\rm obs}$ 
comes from the lower limit $r_{\rm ct}$ (equation \ref{r_ct}).
In the non-dipole field dominated case at the emission region (right panels in figure \ref{region}), 
we fix the model parameters $\alpha_0=\zeta_B=1$. 
The allowed area is larger than that in the dipole case (middle panels in figure \ref{region}). 
The main difference between dipole and non-dipole dominated cases 
is the pitch angle $\alpha$ which could be an order of unity in non-dipole dominated case,  
while the pitch angle near the surface is typically an order of $\alpha\sim10^{-2}$ in the dipole case 
(equation \ref{alpha}).
The efficiency of the synchrotron radiation becomes the maximum value at the lower limit $r=r_{\rm ct}$ 
for both dipole and non-dipole cases.
In the X-ray band ($h\nu_{\rm obs}=1$ keV), 
if the surface dipole field is $B_{\rm s}\lesssim10^{11}$ G,
the lower limit on the emission region is determined by the stellar radius $R_{\rm ns}$ 
(upper right panel in figure \ref{region}). 
In the optical band ($h\nu_{\rm obs}=1~{\rm eV}$), 
the lower limit $r_{\rm ct}$ reaches to the light cylinder radius $R_{\rm lc}$ 
for pulsars with very high spin-down luminosity ($L_{\rm sd}\gtrsim10^{40}$ erg s$^{-1}$).
Then, the optical synchrotron radiation from the magnetosphere is not expected 
for such energetic pulsars irrespective of the existence of the non-dipole component.

 \begin{figure*}
  \begin{center}
   \includegraphics[width=110mm, angle=270]{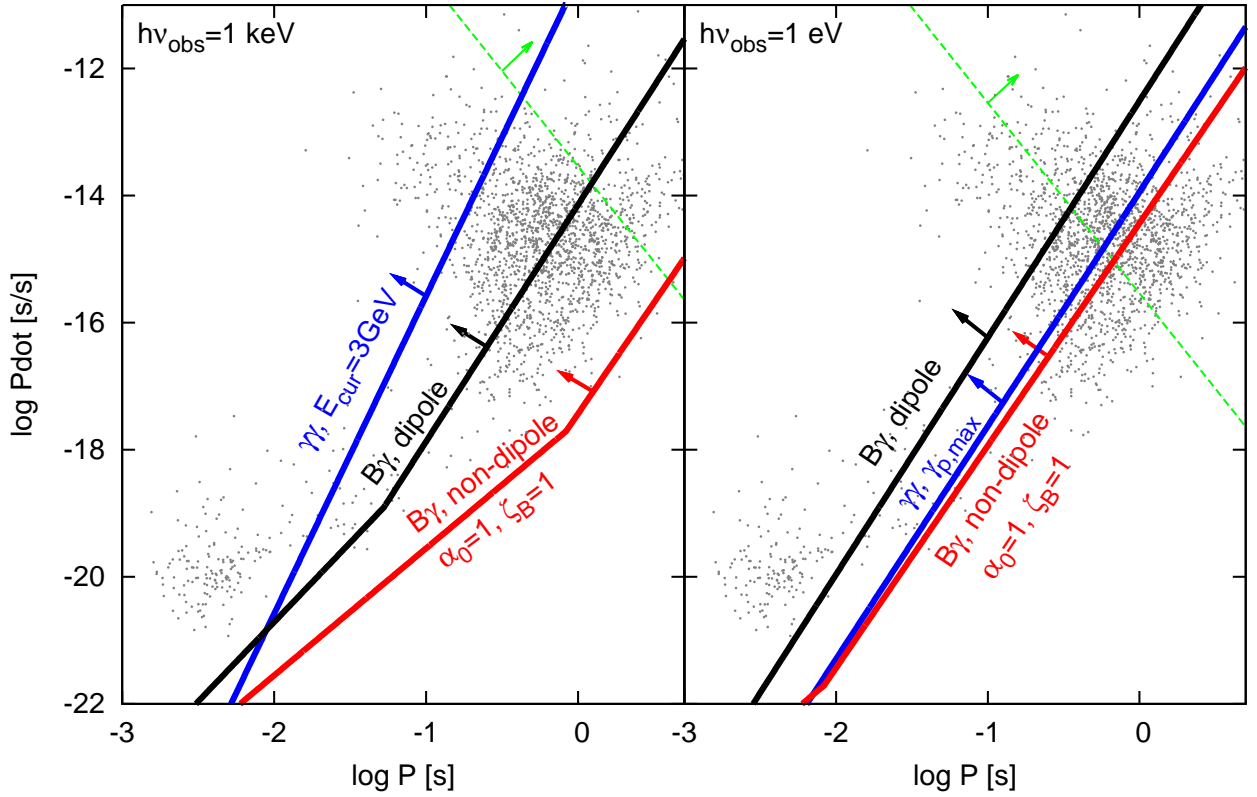}
   \caption{Death lines for the synchrotron radiation with $h\nu_{\rm obs}=1$ keV (left panel), 
and 1 eV (right panel) on the $P$-$\dot{P}$ diagram. 
Blue, black, and red solid lines denote the death lines from equations (\ref{death-line}) 
and (\ref{death-line3}) for $\gamma\gamma$ scenario with $E_{\rm cur}=3$ GeV, 
B$\gamma$ scenario with dipole field, and B$\gamma$ scenario with non-dipole field 
($\alpha_0=\zeta_B=1$), respectively. 
Green dashed lines denote the steady emission conditions with $\epsilon_{\rm B}=10^{-3}$ (left) 
and $10^{-5}$ (right) from equation (\ref{steady-condition}). 
Small dots denote pulsars taken from ATNF Pulsar Catalog \citep{Ma05}. 
Note that in optical case, the death lines for $\gamma\gamma$ scenario are determined by $\gamma_{\rm p,max}$ 
even if $E_{\rm cur}=3$ GeV.}
   \label{p-pdot}
  \end{center}
 \end{figure*}

In figure \ref{p-pdot}, the death lines in $\gamma\gamma$ scenario are shown as blue lines on $P-\dot{P}$ diagram.
We define the death line as the upper limit on the emission region is smaller than the lower limit.  
For pulsars which locate below the death lines in $P$-$\dot{P}$ diagram, 
there is no allowed region for synchrotron radiation from created secondary pairs in their magnetosphere.
Using the equations (\ref{r_gammasyn}) and (\ref{r_gammagamma}),
we derive the death lines for the synchrotron radiation from the pulsar magnetosphere in $\gamma\gamma$ scenario,  
\begin{eqnarray}\label{death-line}
\dot{P}\gtrsim\left\{ \begin{array}{ll}
5.2\times10^{-7}\nu_{\rm obs, keV}^2E_{\rm cur,GeV}^{-9}T_{\rm pc,6.5}^{-5}P_0^5~{\rm s~s}^{-1} & \\ 
~~~~~~~~~ (r_{\gamma\gamma}<r_{\gamma{\rm syn}}, \gamma_{\rm p,E}), & \\
 & \\
1.9\times10^{-14}\nu_{\rm obs, keV}^{1/14}T_{\rm pc,6.5}^{-1/2}P_0^{26/7}~{\rm s~s}^{-1}  & \\ 
~~~~~~~~~ (r_{\gamma\gamma}<r_{\gamma{\rm syn}}, \gamma_{\rm p,max}), & \\
 & \\
1.1\times10^{-14}T_{\rm pc,6.5}^{-2/3}P_0^{11/3}~~{\rm s~s}^{-1}  & \\
~~~~~~~~~ (r_{\gamma\gamma}<R_{\rm lc}, \gamma_{\rm p,max}). & \\
\end{array} \right.
\end{eqnarray}
Condition $r_{\gamma\gamma}<R_{\rm lc}$ in $\gamma\gamma$ scenario with the Lorentz factor of 
primary particles $\gamma_{\rm p,E}$ does not depend on the radius $r$. 
This condition corresponds to the inequality, 
\begin{eqnarray}\label{death-line2}
E_{\rm cur,GeV}T_{\rm pc,6.5}\gtrsim1.4 ~~~~ (r_{\gamma\gamma}<R_{\rm lc}, \gamma_{\rm p,E}), 
\end{eqnarray}
which also does not depend on $P$ and $\dot{P}$. 
From condition (\ref{death-line2}), 
since the observed temperature is typically $T_{\rm pc}\sim10^6$-$10^{6.5}$ K \citep[e.g., ][]{HR93, B09}, 
the characteristic energy 
$E_{\rm cur}\gtrsim$1-3 GeV is required in our model.  
In $\gamma\gamma$ scenario with the Lorentz factor $\gamma_{\rm p,E}$, 
the lower limit on the period derivative $\dot{P}$ depends on $\nu_{\rm obs}^2$ from inequality (\ref{death-line}). 
Then, the allowed parameter area in the $P$-$\dot{P}$ diagram becomes large for the low frequency $\nu_{\rm obs}$.
However, since the Lorentz factor $\gamma_{\rm p,E}$ cannot exceed $\gamma_{\rm p,max}$, 
we have to use $\gamma_{\rm p,max}$ to derive the death line in the case $\gamma_{\rm p,E}>\gamma_{\rm p,max}$. 
In the X-ray band, we consider the death line with $\gamma_{\rm p,E}<\gamma_{\rm p,max}$ 
in most region on $P$-$\dot{P}$ diagram while in the optical band, we should consider the case 
$\gamma_{\rm p,E}>\gamma_{\rm p,max}$ for the death line.
In the latter case, the death line is almost insensitive to $\nu_{\rm obs}$ (inequality \ref{death-line}). 
Therefore, we do not expect any synchrotron radiation with frequency $h\nu_{\rm obs}\lesssim1{\rm eV}$ from pulsars
which reside below the death line with $\gamma_{\rm p,max}$ in $\gamma\gamma$ scenario.

In B$\gamma$ scenario, the death lines for the synchrotron radiation are derived from equations (\ref{r_Bgamma}) and 
(\ref{r_ct}) as,
\begin{eqnarray}\label{death-line3}
\dot{P}\gtrsim\left\{ \begin{array}{ll}
7.2\times10^{-15}\nu_{\rm obs,keV}^{-6/11}P_0^{41/11}~~{\rm s~s}^{-1} & \\ 
~~~~~~~~~ (r_{\rm ct}<r_{\rm B{\gamma}}, {\rm dipole}), & \\
 & \\
1.9\times10^{-16}P_0^{5/2}~~{\rm s~s}^{-1} & \\ 
~~~~~~~~~ (R_{\rm ns}<r_{\rm B\gamma}, {\rm dipole}), & \\
 & \\
3.7\times10^{-18}\alpha_0^{-7/4}\zeta_{\rm B}^{1/4}\nu_{\rm obs,keV}^{-1}P_0^{7/2}~~{\rm s~s}^{-1} & \\ 
~~~~~~~~~ (r_{\rm ct}<r_{\rm B{\gamma}}, {\rm non-dipole}), & \\
 & \\
2.8\times10^{-18}\alpha_0^{-1/2}\zeta_{\rm B}^{-1/2}P_0^2~~{\rm s~s}^{-1} & \\ 
~~~~~~~~~ (R_{\rm ns}<r_{\rm B\gamma}, {\rm non-dipole}). & \\
\end{array} \right.
\end{eqnarray}
We also show these death lines as black (dipole) and red lines (non-dipole) in figure \ref{p-pdot}. 
The dependence on the observed frequency $\nu_{\rm obs}$ 
mainly comes from the condition of the validity of the synchrotron approximation 
(inequality \ref{frequency_condition}).
From the left panel of figure \ref{p-pdot}, 
we expect that the synchrotron radiation with the frequency $h\nu_{\rm obs}>1$ keV 
is emitted from the magnetosphere of almost all detected radio pulsars 
if the non-dipole component of the magnetic field dominates at the emission region. 
On the other hand, in the optical band ($h\nu_{\rm obs}\sim1~{\rm eV}$), 
we do not expect the emission of the synchrotron radiation from pulsars with $L_{\rm sd}\lesssim10^{32}$ erg s$^{-1}$. 
Even if the non-dipole component enhances the magnetic field at the emission region $(\zeta_{\rm B}>1)$, 
the area of the allowed region with larger $P$ ($P\gtrsim1$ s for X-ray, and $P\gtrsim10^{-2}$ s for optical) 
on the $P$-$\dot{P}$ diagram decreases because the lower limit $r_{\rm ct}$ increases. 

\subsection{Maximum Luminosity of Synchrotron Radiation}
\label{sec:maximum}

 \begin{figure*}
  \begin{center}
   \includegraphics[width=90mm, angle=270]{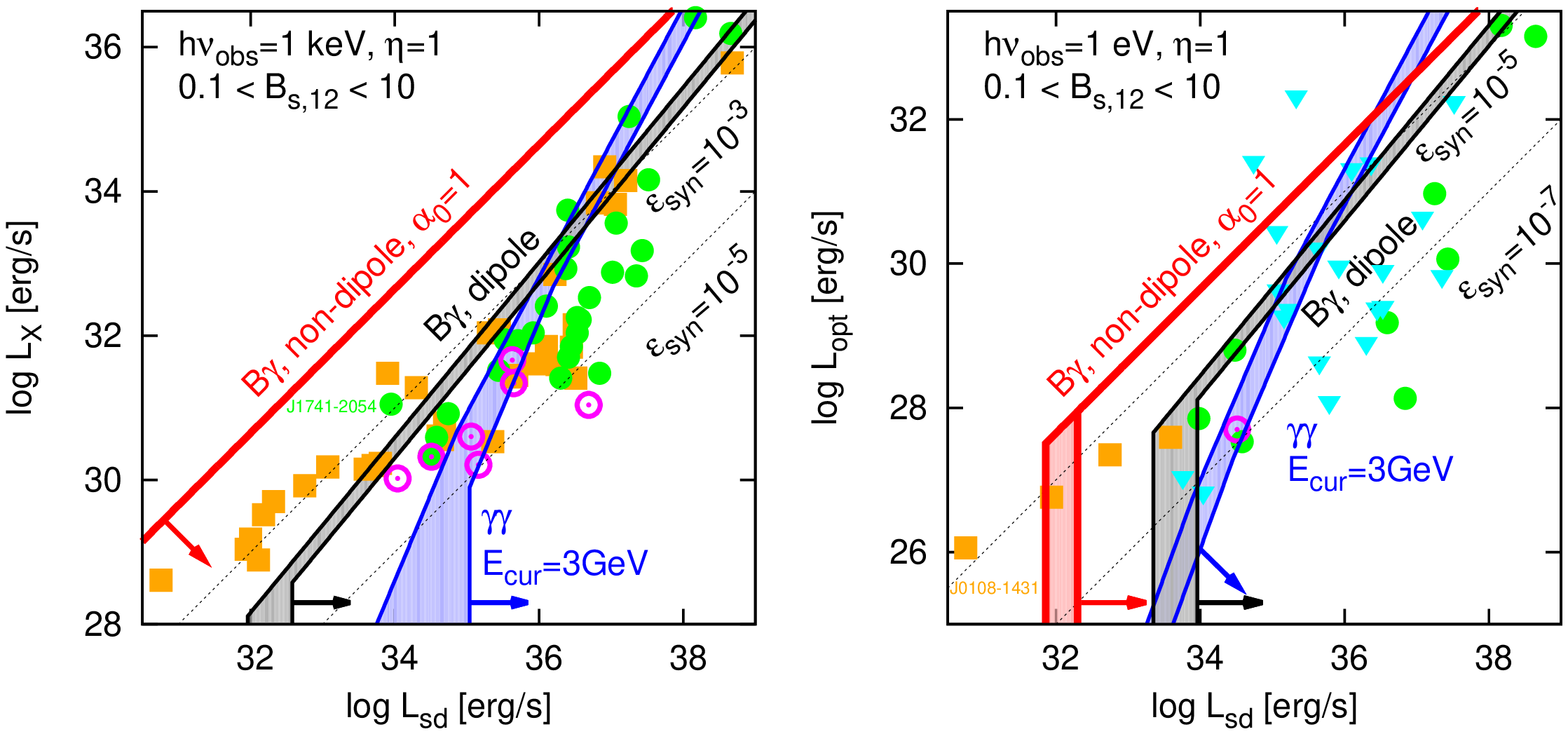}
   \caption{
The maximum luminosities of synchrotron radiation in X-ray (left) and optical bands (right) 
for scenarios of $\gamma\gamma$ with $E_{\rm cur}=3$ GeV (blue), B$\gamma$ with dipole field (black), 
and B$\gamma$ with non-dipole field (red) as a function of $L_{\rm sd}$. 
The observed luminosities in X-ray and optical bands are also plotted. 
Data are taken from \citet{KP08, Kar12, Pos12, Ab13, PB15, Sza+17} in X-ray, 
and from \citet{ZM13, Ber15, Kir15, Mig16a, Mig16b, Mig16c, Shi16} in optical. 
Filled green and open magenta circles are radio-loud and radio-quiet $\gamma$-ray pulsars, respectively. 
Orange squares denote the non-$\gamma$-ray pulsars. 
In the right panel, upper limits on the optical luminosity are also plotted as cyan triangles. 
The parameters are fixed to $\alpha_0=\eta=\zeta_{\rm B}=1$. Note that the dependence of $\zeta_{\rm B}$ 
only appears to the lower boundary of $L_{\rm sd}$ in B$\gamma$ scenario with non-dipole field.}
   \label{efficiency}
  \end{center}
 \end{figure*}

Using equations (\ref{epsilon_syn,gammagamma}) and (\ref{epsilon_syn,Bgamma}), 
and taking into account four constraints on the emission region (\ref{r_gammasyn} - \ref{r_gammagamma}), 
we calculate the maximum luminosity of synchrotron radiation as a function of model parameters 
$\eta$, $\alpha_0$ and $\zeta_{\rm B}$. 
In $\gamma\gamma$ scenario, the synchrotron luminosity becomes maximum at $r=R_{\rm lc}$ (equation \ref{R_lc}) 
or $r_{\gamma{\rm syn}}$ 
(equation \ref{r_gammasyn}), 
$r=r_{\rm eq,syn}$ (equation \ref{r_eq,syn}), and $r=r_{\gamma\gamma}$  (equation \ref{r_gammagamma}) 
for panels A, B and C in figure \ref{schematics}, respectively. 
Substituting these radius into equation (\ref{epsilon_syn,gammagamma}), we obtain the maximum luminosity.
For the low-$L_{\rm sd}$ pulsars which satisfy condition $t_{\rm cool, syn}>t_{\rm ad}$ at the emission region 
(pane C), 
Lorentz factor of the primary particles could become 
$\min\{\gamma_{\rm p,max}, \gamma_{\rm p,E}\}=\gamma_{\rm p,max}$. 
We use $\gamma_{\rm p,max}$ only for the the condition in panel C. 
The synchrotron luminosities are described as,
\begin{eqnarray}\label{L_syn,gammagamma} 
L_{\rm syn}\lesssim\left\{ \begin{array}{ll} 
4.6\times10^{23}\eta\nu_{\rm obs,keV}^{1/2}E_{\rm cur,GeV}^{-1} & \\ 
~~~\times\epsilon_{\rm pc,-3}T_{\rm pc,6.5}^{-1}B_{\rm s,12}^{-1/4}L_{\rm sd,31}^{15/8}~~{\rm erg~s}^{-1}  & \\ 
~~~~~~~ (A, r_{\gamma{\rm syn}}>R_{\rm lc}, \gamma_{\rm p,E}) & \\ 
 & \\ 
4.1\times10^{20}\eta\nu_{\rm obs,keV}^{-1/5}E_{\rm cur,GeV}^{2/5} & \\ 
~~~\times\epsilon_{\rm pc,-3}T_{\rm pc,6.5}^{-1}B_{\rm s,12}^{-3/5}L_{\rm sd,31}^{12/5}~~{\rm erg~s}^{-1}  & \\ 
~~~~~~~ (A, r_{\gamma{\rm syn}}<R_{\rm lc}, \gamma_{\rm p,E}) & \\ 
 & \\ 
2.3\times10^{22}\eta\nu_{\rm obs,keV}^{10/13}E_{\rm cur,GeV}^{-1} & \\ 
~~~\times\epsilon_{\rm pc,-3}T_{\rm pc,6.5}^{-1}B_{\rm s,12}^{-5/13}L_{\rm sd,31}^{61/26}~~{\rm erg~s}^{-1} & \\ 
~~~~~~~ (B, \gamma_{\rm p,E}) & \\
 & \\
1.1\times10^{21}\eta\nu_{\rm obs,keV}E_{\rm cur,GeV}^{1/2} & \\ 
~~~\times\epsilon_{\rm pc,-3}T_{\rm pc,6.5}^{1/2}B_{\rm s,12}^{-1/2}L_{\rm sd,31}^{11/4}~~{\rm erg~s}^{-1}  & \\
~~~~~~~ (C, \gamma_{\rm p,E}) & \\
 & \\
4.0\times10^{18}\eta\nu_{\rm obs,keV} & \\ 
~~~\times\epsilon_{\rm pc,-3}T_{\rm pc,6.5}B_{\rm s,12}^{-1}L_{\rm sd,31}^{9/2}~~{\rm erg~s}^{-1}  & \\
~~~~~~~ (C, \gamma_{\rm p,max}). & \\
\end{array} \right. 
\end{eqnarray}
In $\gamma\gamma$ scenario, a model parameter is only $\eta$. 
Note that if we extrapolate the derived luminosity to highly energetic pulsars 
($L_{\rm sd}\gtrsim10^{40}$ erg s$^{-1}$) in a simplistic form, 
the luminosity of the synchrotron radiation with $\eta=1$ seems to exceed the spin-down luminosity $L_{\rm sd}$.
However, the optical depth $\tau_{\gamma\gamma}$ becomes higher than unity for such energetic pulsars. 
Then, the number flux of the created pair does not depend on the optical depth 
$(\min\{\tau_{\gamma\gamma},1\}=1$ in equation \ref{dotN_s}).
As a result, the luminosity of the synchrotron radiation 
follows the relation $L_{\rm syn}\propto L_{\rm sd}^{5/8}$, which always satisfies $L_{\rm syn}<L_{\rm sd}$ 
even if $\eta=1$. 

On the other hand, the maximum luminosity of synchrotron radiation in B$\gamma$ scenario are derived as,
\begin{eqnarray}\label{L_syn,Bgamma}
L_{\rm syn}\lesssim\left\{ \begin{array}{ll} 
6.4\times10^{26}\eta\nu_{\rm obs,keV}^{5/7} & \\ 
~~~\times B_{\rm s,12}^{-1/7}L_{\rm sd,31}^{17/14}~~{\rm erg~s}^{-1}  & \\
~~~~~~~ (D, {\rm dipole}), & \\
 & \\
4.5\times10^{29}\alpha_0^2\eta\nu_{\rm obs,keV}L_{\rm sd,31}~~{\rm erg~s}^{-1}  & \\
~~~~~~~ (E, {\rm non-dipole}). & \\
\end{array} \right. 
\end{eqnarray}
The maximum luminosities are given at $r=r_{\rm ct}$. 
In the non-dipole case, the lower limit depends on the magnetic field, 
$r_{\rm ct}\propto\zeta_{\rm B}^{1/3}$ (equation \ref{r_ct}). 
The dependence of the luminosity $L_{\rm syn}$ on the parameter $\zeta_{\rm B}$ and the emission region $r$ is 
$L_{\rm syn}\propto \zeta_{\rm B}^{1/2}r^{-3/2}$ (equation \ref{epsilon_syn,Bgamma}). 
Then, the maximum luminosity in the non-dipole field dominant case does not depend on the parameter $\zeta_{\rm B}$.

In figure \ref{efficiency}, we show the derived upper limits with $\alpha_0=\eta=1$ 
on the $L_{\rm syn}$ versus $L_{\rm sd}$ planes. 
For the range of the surface dipole field, we consider $10^{11}{\rm G}<B_{\rm s}<10^{13}$G. 
We plot the observed luminosities of the non-thermal X-ray $L_{\rm X}$ (left panel) and 
optical emissions $L_{\rm opt}$ (right panel), and the spin-down luminosities $L_{\rm sd}$. 
For the non-thermal optical emission, we also plot the observed upper limits.

In the X-ray band, 
the observed non-thermal luminosities of most pulsars with $L_{\rm sd}\gtrsim10^{35}$ erg s$^{-1}$ 
are comparable to or lower than the maximum luminosity with $\eta=1$ 
in $\gamma\gamma$ pair creation scenario 
(equation \ref{L_syn,gammagamma}; blue curve in the left panel of figure \ref{efficiency}). 
The observed X-ray luminosities 
are $\sim10^{-2}-1$ times lower than the maximum value. 
Then, the conversion efficiency should be $\eta\gtrsim10^{-2}$. 
On the other hand, for pulsars with $L_{\rm sd}\lesssim10^{35}$ erg s$^{-1}$ including some $\gamma$-ray pulsars 
(green and magenta points), the observed non-thermal luminosities exceed the maximum luminosity in $\gamma\gamma$ scenario. 
These observed X-ray luminosities are explained 
in B$\gamma$ scenario with dipole magnetic field (equation \ref{L_syn,Bgamma}; black curve), 
except for PSR J1741-2054. 
For J1741-2054 and other pulsars with $L_{\rm sd}\lesssim10^{34}$ erg s$^{-1}$, 
we should take into account the non-dipole component of the magnetic field at the emission region 
(equation \ref{L_syn,Bgamma}; red line). 
Then, the observed luminosities for all samples are lower than the maximum value with $\eta=1$ and $\alpha_0=1$.

In the optical band, the right panel of figure \ref{efficiency} shows that 
the luminosity of the synchrotron radiation with $\eta=1$ in $\gamma\gamma$ scenario 
(equation \ref{L_syn,gammagamma}; blue curve) is also higher than the observed non-thermal optical luminosities 
for pulsars with $L_{\rm sd}\gtrsim10^{35}$ erg s$^{-1}$. 
For pulsars with $10^{33}$ erg s$^{-1}\lesssim L_{\rm sd}\lesssim10^{35}$ erg s$^{-1}$, 
the observed optical luminosities are lower than the maximum luminosity with $\eta=1$ in B$\gamma$ scenario 
with dipole magnetic field (equation \ref{L_syn,Bgamma}; black curve).
For pulsars with $10^{32}$ erg s$^{-1}\lesssim L_{\rm sd}\lesssim10^{33}$ erg s$^{-1}$, 
the non-dipole component are required for which the maximum luminosity with $\eta=1$ and $\alpha_0=1$ 
exceeds their observed non-thermal optical luminosities  (equation \ref{L_syn,Bgamma}; red curve).
For pulsars with $L_{\rm sd}\lesssim10^{32}$ erg s$^{-1}$, 
from condition $r_{\rm ct}<r_{\rm B\gamma}$ (equation \ref{death-line3}), 
the synchrotron radiation in our model is not expected in optical band. 
The lower limit on the spin-down luminosity is proportional to 
the model parameter, $L_{\rm sd}\propto\zeta_{\rm B}^{1/4}$, derived from inequality (\ref{death-line3}), 
so that the strong magnetic field $\zeta_{\rm B}>1$ increases the lower limit on $L_{\rm sd}$.
PSR J0108-1431 with $L_{\rm sd}\sim8\times10^{30}$ erg s$^{-1}$ have been detected in optical band \citep{MPK08}. 
We will discuss this object in section \ref{sec:discussion}.

\subsection{Luminosity Ratio}
\label{sec:ratio}

 \begin{figure*}
  \begin{center}
   \includegraphics[width=90mm, angle=270]{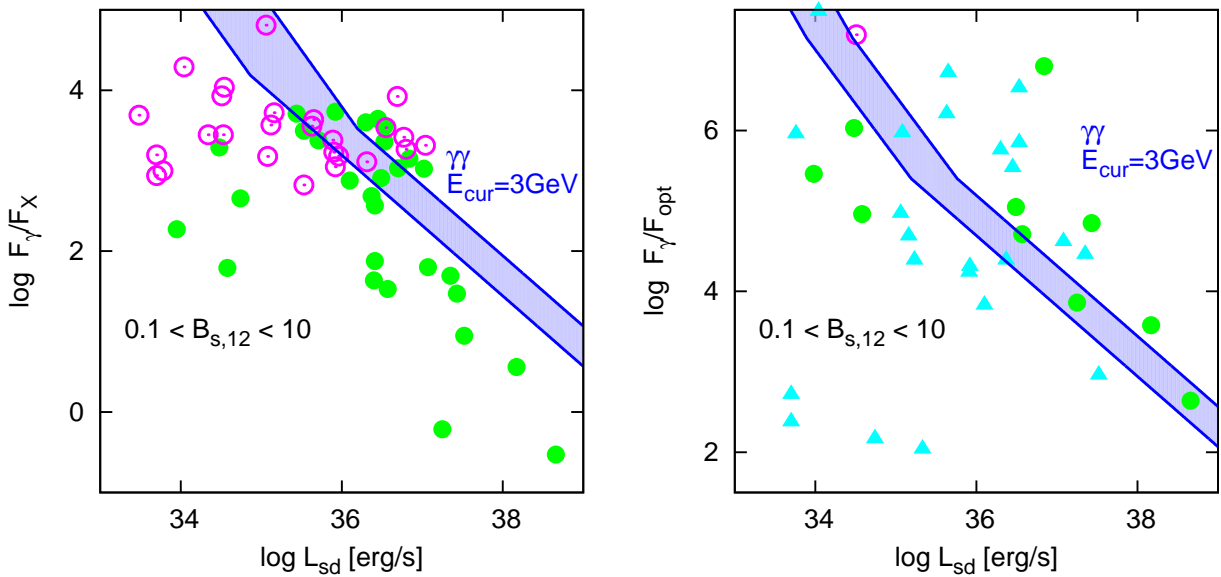}
   \caption{
Flux ratios $F_{\gamma}/F_{\rm X}$ (left) and $F_{\gamma}/F_{\rm opt}$ (right) versus $L_{\rm sd}$ for $\gamma$-ray pulsars. 
The symbols are the same as in figure \ref{efficiency}. 
Note that the cyan triangles are lower limits on $F_{\gamma}/F_{\rm opt}$ in the right panel.
Data are taken from \citet{Ab13, Mar15, Ber15}.
Blue curves are the ratio $L_{\rm cur}/L_{\rm syn}$ in $\gamma\gamma$ scenario, which is calculated from equations 
(\ref{L_syn,gammagamma}) and (\ref{L_cur,full}). 
}
   \label{ratio}
  \end{center}
 \end{figure*}

In our model, the fluxes of the curvature and the synchrotron radiations should be related. 
Then, the observed $\gamma$-ray, X-ray and optical emissions should also be related, 
unless the curvature photons are significantly absorbed in the magnetosphere. 
In $\gamma\gamma$ scenario, the absorption of the curvature photons is negligible 
for most $\gamma$-ray pulsars because of the low optical depth $\tau_{\gamma\gamma}\ll1$.  
The luminosity of curvature radiation in our model is described by, 
\begin{eqnarray}\label{L_cur,full}
L_{\rm cur}\sim P_{\rm cur}\dot{N}_{\rm p}\min\{t_{\rm cool,cur},t_{\rm ad}\}.
\end{eqnarray} 
Within the typical range of the parameters for $\gamma$-ray pulsars, 
condition $t_{\rm cool,cur}<t_{\rm ad}$ is satisfied (equations \ref{r_eq,cur-E_cur} and \ref{L_eq,cur}).
Using this condition and equations (\ref{dotN_p}) and (\ref{t_cool,cur}), 
the curvature luminosity $L_{\rm cur}$ is derived as
\begin{eqnarray}\label{L_cur}
L_{\rm cur}\sim\eta L_{\rm sd}. 
\end{eqnarray}
From equation (\ref{epsilon_syn,gammagamma}), the synchrotron efficiency is 
proportional to the conversion efficiency, $\epsilon_{\rm syn}\propto\eta$. 
Then, the luminosity (or the flux) ratio $L_{\rm cur}/L_{\rm syn}\sim(\epsilon_{\rm syn}/\eta)^{-1}$ 
does not depend on the conversion efficiency $\eta$. 

In figure \ref{ratio}, we plot the observed $\gamma$-ray to X-ray (left panel) and $\gamma$-ray to optical 
(right panel) flux ratios for $\gamma$-ray pulsars. 
Since the flux ratios do not depend on the distance from us, 
we also plot the observed ratios for the radio-quiet $\gamma$-ray pulsars (magenta) 
whose distance are not well constrained. 
For the pulsars with $L_{\rm sd}\gtrsim10^{36}$ erg s$^{-1}$, the observed flux ratios are an order of magnitude 
smaller than our model for X-ray band. 
As discussed in section \ref{sec:fluxdiscussion}, 
for energetic pulsars ($L_{\rm sd}\gtrsim10^{36}$ erg s$^{-1}$),
the thermal photons from entire surface which we do not consider in the model 
could contribute to the optical depth $\tau_{\gamma\gamma}$.
The observed flux ratios of pulsars with $L_{\rm sd}\lesssim10^{35}$ erg s$^{-1}$ are also 
much smaller than our model for both X-ray and optical bands. 
For these low-$L_{\rm sd}$ pulsars, we should consider other effects such as the contribution of 
the synchrotron radiation from inner region in B$\gamma$ scenario as discussed in section \ref{sec:discussion}.
Note that in B$\gamma$ scenario, significant curvature photons could be absorbed in the magnetosphere.
Then, the observed $\gamma$-ray flux could be much smaller than X-ray and optical fluxes, 
so that we do not plot the flux ratio in B$\gamma$ scenario in figure \ref{ratio}. 

\section{DISCUSSION}
\label{sec:discussion}

In this paper, we analytically calculate the luminosity of the synchrotron radiation 
from the secondary particles created in the pulsar magnetosphere. 
In order to constrain the efficiency of the energy conversion from the loss rate of the rotation energy to 
the kinetic energy flux of the particles, we introduce a model parameter $\eta$ and compare with observations 
in X-ray and optical bands. 
X-ray and optical emissions are detected in not only young, high-$L_{\rm sd}$ 
$\gamma$-ray pulsars ($\tau_{\rm c}\lesssim10^6$ yr, $L_{\rm sd}\gtrsim10^{33}$-$10^{34}$ erg s$^{-1}$), 
but also old, low-$L_{\rm sd}$ pulsars ($\tau_{\rm c}\gtrsim10^6$ yr, $L_{\rm sd}\lesssim10^{33}$-$10^{34}$ erg s$^{-1}$).
We consider $\gamma\gamma$ and B$\gamma$ pair creation processes
as the electron/positron pair conversion process from the curvature photons emitted by the primary particles 
(figure \ref{image}). 
For the energy of primary particles, we use the observed value of the $\gamma$-ray spectral cutoff energy 
in $\gamma\gamma$ scenario. 
This model does not need to assume the strength of the electric field at the particle acceleration region, 
which is highly uncertain.  
In B$\gamma$ scenario, we use the maximum value of the potential drop across the polar cap to obtain the energy 
of the primary particles, although the derived synchrotron luminosity does not depend on the Lorentz factor of 
the primary particles.
The newly created secondary particles have a non-zero value of a pitch angle and emit the synchrotron radiation.
For the magnetic field, we assume the dipole field in the magnetosphere. 
Near the stellar surface, we also include the effects of the non-dipole field on the pitch angle of
the secondary particles and the curvature radius of the magnetic field line.
In order to calculate the maximum synchrotron luminosity, 
we take into account the requirements for the radiation mechanism being synchrotron radiation.

In B$\gamma$ scenario, the region where the luminosity of the synchrotron radiation becomes maximum
is near the neutron star surface (figure \ref{region}). 
The synchrotron radiation from secondary pairs can be detected from pulsars with the spin-down luminosity 
$L_{\rm sd}\gtrsim10^{32}$ erg s$^{-1}$ for X-ray band and $L_{\rm sd}\gtrsim10^{33}$ erg s$^{-1}$ 
for optical band in our B$\gamma$ scenario with the dipole magnetic field (figure \ref{p-pdot}).  
The synchrotron efficiency for typical pulsars is $\epsilon_{\rm syn}\sim(10^{-4}-10^{-3})\nu_{\rm obs,keV}^{5/7}$. 
The dependence on the spin-down luminosity is $L_{\rm syn}\propto L_{\rm sd}^{17/14}$ which is close to the linear trend. 
While the observed luminosities for high-$L_{\rm sd}$ pulsars ($L_{\rm sd}\gtrsim10^{34}$ erg s$^{-1}$) 
could be explained by our model with dipole dominated case, 
the luminosities for low-$L_{\rm sd}$ pulsars ($L_{\rm sd}\lesssim10^{34}$ erg s$^{-1}$) 
exceed the maximum synchrotron luminosity with $\eta=1$. 
In the non-dipole dominant case at the emission region, 
most pulsars reside the allowed area on the $P$-$\dot{P}$ diagram for X-ray band (figure \ref{efficiency}).
On the other hand, we expect that the synchrotron radiation is not detected from pulsars with 
$L_{\rm sd}\lesssim10^{32}$ erg s$^{-1}$ in the optical band. 
The luminosity of the synchrotron radiation is $L_{\rm syn}\sim4\times10^{-2}\alpha_0^2\eta\nu_{\rm obs,keV}L_{\rm sd}$, 
which does not depend on the magnetic field strength at the emission region. 

In $\gamma\gamma$ scenario (figure \ref{efficiency}), 
the allowed region of the synchrotron radiation resides near the light cylinder (figure \ref{region}).
The synchrotron death line in X-ray band corresponds to $L_{\rm sd}\sim10^{34}-10^{35}$ erg s$^{-1}$.
In optical band, the synchrotron death line corresponds to $L_{\rm sd}\sim10^{32}$ erg s$^{-1}$, 
although the synchrotron efficiency becomes much small ($\epsilon_{\rm syn}\lesssim10^{-7}\eta$) 
for the range $L_{\rm sd}\lesssim10^{34}$ erg s$^{-1}$.
The synchrotron luminosity dependence on the spin-down luminosity is $L_{\rm syn}\propto L_{\rm sd}^{a}$ with $a\sim2$. 
In our model, the normalized thermal luminosity from the heated polar cap $\epsilon_{\rm pc}$ is constant 
based on the observations \citep{BT97, B09}. 
Then, the number of thermal photons is proportional to the spin-down luminosity.
The synchrotron luminosity is also proportional to the total energy of the primary particles, 
so that the index becomes $a\sim2$.
The observed non-thermal X-ray and optical emissions for $\gamma$-ray pulsars 
with $L_{\rm sd}\gtrsim10^{35}$ erg s$^{-1}$ are lower than the maximum luminosity of our model with $\eta=1$ 
(figure \ref{efficiency}). 
On the other hand, for the $\gamma$-ray pulsars with low spin-down luminosity, $L_{\rm sd}\lesssim10^{35}$ erg s$^{-1}$, 
their observed luminosities exceed the maximum luminosity with $\eta=1$ in $\gamma\gamma$ scenario. 
Since $\gamma$-ray pulsars show no significant absorption feature via B$\gamma$ process 
in their observed spectra \citep{Ab13}, 
the curvature radiation by the primary particles should be emitted from the outer magnetosphere. 
Then, in addition to the outer magnetosphere, another synchrotron emitting region 
may reside for $\gamma$-ray pulsars (see section \ref{sec:fluxdiscussion}).

Although our model only consider the region $r\le R_{\rm lc}$, 
the derived synchrotron luminosity in $\gamma\gamma$ scenario with $r=R_{\rm lc}$ 
gives the maximum value for the cases 
where particle acceleration and high-energy emission take place in the current sheet outside the light cylinder. 
Recent numerical simulations indicate that 
a significant fraction of Poynting flux dissipates at $\sim1-2R_{\rm lc}$ \citep{B15, CPPS15}. 
Since the dissipation region is very close to the light cylinder ($\lesssim2R_{\rm lc}$), 
the synchrotron efficiency would be comparable to the values in equation (\ref{epsilon_syn,gammagamma}) 
with $r=R_{\rm lc}$. 
If our model extend outside the light cylinder, 
the maximum synchrotron efficiency could be given at $r=R_{\rm lc}$ as follows. 
The synchrotron efficiency in equation (\ref{epsilon_syn,gamma-gamma}) 
is applicable to the region outside the light cylinder if the conditions for the timescales, 
$t_{\rm cool,syn}<t_{\rm ad}$ and $t_{\rm cool,cur}<t_{\rm ad}$, are satisfied. 
For simplicity, we assume $\alpha\sim1$, $\cos\theta_{\rm col}\sim0$, and $R_{\rm cur}\sim r$ 
at the region $r>R_{\rm lc}$. 
Since the toroidal component of the magnetic field would be dominant outside the light cylinder, 
we assume $B\propto r^{-1}$ at $r>R_{\rm lc}$. 
Then, the Lorentz factor of synchrotron emitting particles and the optical depth are 
$\gamma_{\rm s,syn}\propto r^{1/2}$ (equation \ref{gamma_s,syn}) and $\tau_{\gamma\gamma}\propto r^{-1}$ 
(equation \ref{tau_gammagamma}), respectively. 
We also consider $\gamma_{\rm p,E}<\gamma_{\rm p,max}$ for the Lorentz factor of the primary particles, 
so that the Lorentz factor $\gamma_{\rm s,pair}$ does not depend on $r$ (equation \ref{gamma_s,pair}). 
Using equation (\ref{epsilon_syn,gamma-gamma}), the synchrotron efficiency deceases as the large distance, 
$\epsilon_{\rm syn}\propto r^{-1/2}$, and the efficiency becomes maximum at the light cylinder radius $r=R_{\rm lc}$. 
Therefore, $\gamma\gamma$ scenario with $r=R_{\rm lc}$ gives the upper limit on the efficiency $\epsilon_{\rm syn}$ 
even if most particle acceleration takes place outside the light cylinder. 

Note that we use $\gamma_{\rm p,E}$ as the Lorentz factor of primary particles. 
Then, the effective number flux of primary particle $\dot{N}_{\rm p}$ is 
$\eta\gamma_{\rm p,max}/\gamma_{\rm p,E}$ times larger than the GJ number flux, 
$2\pi r_{\rm pc}^2\rho_{\rm GJ, sur}c/e$ \citep{GJ69}, 
where $\rho_{\rm GJ, sur}$ is GJ charge density at the neutron star surface and $r_{\rm pc}$ 
is the polar cap radius. 
The primary particle is a carrier of current flowing the magnetosphere. 
It seems that the averaged current density exceeds the GJ value $\rho_{\rm GJ}c$. 
However, if the cooling timescale is smaller than the advection timescale, 
the primary particles are continuously accelerated and lose their energy via curvature radiation in the acceleration region. 
Then, the actual number of the particles could be $t_{\rm cool, cur}/t_{\rm ad}$ times smaller than the effective value. 
The minimum number of the primary particles normalized by the GJ value is 
$\eta(\gamma_{\rm p,max}/\gamma_{\rm p,E})(t_{\rm cool, cur}/t_{\rm ad})\sim1.1\eta(E_{\rm cur}/3{\rm GeV})^{-4/3}(B_{\rm s}/10^{12}{\rm G})^{-1/6}(L_{\rm sd}/10^{35}{\rm erg~s}^{-1})^{7/12}$, which is an order of unity.
Therefore, in our model, we do not consider that the current flowing the magnetosphere 
significantly exceeds the GJ current. 

The synchrotron luminosities with $\eta=1$ (equations \ref{L_syn,gammagamma} and \ref{L_syn,Bgamma}) provide the 
maximum luminosities which correspond to the theoretical critical lines on 
$L_{\rm X}$ versus $L_{\rm sd}$ and $L_{\rm opt}$ versus $L_{\rm sd}$ planes. 
Using the observed non-thermal X-ray luminosity, 
\citet{Kar12} found the critical lines which are described by broken power-law functions with the transition point 
$L_{\rm sd}\sim10^{35}$ erg s$^{-1}$. 
Their transition point is consistent with the point 
where the synchrotron luminosities of $\gamma\gamma$ and B$\gamma$ scenarios are comparable in our model. 
For $L_{\rm sd}\gtrsim10^{35}$ erg s$^{-1}$, the critical line suggested by \citet{Kar12} 
is roughly consistent with the maximum luminosity in $\gamma\gamma$ scenario. 
For $10^{33}$ erg s$^{-1}\lesssim L_{\rm sd}\lesssim10^{35}$ erg s$^{-1}$, the index of their derived critical line is small, 
$L_{\rm X}\propto L_{\rm sd}^{0.38}$ \citep{Kar12}. 
The small value of the index is possible if pulsars which have the non-dipole field at the emission region are 
preferentially detected at the range $L_{\rm sd}\lesssim10^{34}$ erg s$^{-1}$.
For $L_{\rm sd}\lesssim10^{33}$ erg s$^{-1}$, 
we suggest that the critical line follows $L_{\rm syn}\propto L_{\rm sd}$ 
from B$\gamma$ scenario with the non-dipole field.
Our theoretical critical lines would be useful to select the observational target in optical and X-ray bands.

For PSR J0108-1431, the observed luminosity of the non-thermal X-ray emission is within the maximum value in our model,
although the extreme conditions that the non-dipole field is dominant 
and the conversion efficiency $\eta$ and the pitch angle $\alpha_0$ are an order of unity are required.
However, the non-thermal optical emission via synchrotron radiation is not expected from PSR J0108-1431 in our model.
Even if we take into account the effect of the non-dipole component of the magnetic field, 
the synchrotron luminosity does not depend on the strength of the magnetic field $\zeta_{\rm B}$.
Instead, the large value of $\zeta_{\rm B}(>1)$ makes the lower boundary of the spin-down luminosity high
because the death line for condition $r_{\rm ct}<r_{\rm B\gamma}$ is $\dot{P}\propto\zeta_{\rm B}^{1/4}$ 
(equation \ref{death-line3}). 
The significance of the optical detection of J0108-1431 is marginal \citep{MPK08} 
and subsequent observation could not detect this pulsar \citep{MPK11}.
If the non-thermal optical emission from J0108-1431 is confirmed, 
we should consider the other emission mechanisms and/or the other energy sources.
Future observations will confirm whether the detected signal comes from J0108-1431 or not. 

The region where the optical luminosity of the synchrotron radiation becomes maximum is $r\sim10-50R_{\rm ns}$ 
in B$\gamma$ scenario (figure \ref{region}).
This region is also considered as the coherent radio emission site from observations \citep[e.g., ][]{Pilia16}.
For low-$L_{\rm sd}$ pulsars ($L_{\rm sd}\lesssim10^{34}$ erg s$^{-1}$) which are required for the non-dipole component 
to explain their X-ray and optical luminosities, 
the radio pulse profile has high complexity \citep{KJ07} and the degree of linear polarization in radio 
is significantly lower than the high $L_{\rm sd}$ pulsars \citep{WJ08}.
The modulation phenomena such as nulling and mode changing are also seen in the low-$L_{\rm sd}$ pulsars \citep{WMJ07}. 
These facts support that the non-dipole magnetic field is a main control parameter for not only the X-ray and optical 
emission but also the radio emission for low-$L_{\rm sd}$ pulsars. 

\subsection{Energy Conversion Efficiency}
\label{sec:efficiencydiscussion}

Comparing with observed values (figure \ref{efficiency}), 
the conversion efficiency $\eta$ should be $\gtrsim0.01-1$ 
in $\gamma\gamma$ scenario for pulsars with $L_{\rm sd}\gtrsim10^{35}$ erg s$^{-1}$. 
In B$\gamma$ scenario with the dipole dominant case, 
the efficiency should be $\eta\gtrsim0.01-1$ for pulsars with $L_{\rm sd}\gtrsim10^{34}$ erg s$^{-1}$. 
Even if we consider the effects of the non-dipole component in B$\gamma$ scenario, the efficiency should be
$\eta\gtrsim10^{-4}$-$0.1$ with $\alpha_0=1$. 
We do not find any dependence of the efficiency $\eta$ on $P$ and $\dot{P}$.
Note that our model has optimistic assumptions to enlarge the synchrotron luminosity as also discussed in \citet{KT14}.
Therefore, the actual efficiency $\eta$ would be larger than that derived from our model.

The efficiency parameter $\eta$ almost corresponds to $\sim(1+\sigma)/\sigma$, 
where $\sigma$ is the magnetization parameter (the ratio of the Poynting to the kinetic energy fluxes).
Although the magnetization parameter is usually considered as $\sigma\gg1$ in the magnetosphere \citep[e.g., ][]{DH82},
the requirement of $\eta\sim O(0.1)$ ($\sigma\sim O(10)$) for observations means that the significant electromagnetic energy have 
to convert to the particle energy in the magnetosphere. 
Especially, in B$\gamma$ scenario, the energy conversion to $\eta\sim O(0.1)$ should occur near the stellar surface. 

In B$\gamma$ scenario, 
the emission regions where the synchrotron luminosities in optical and X-ray become maximum are different 
(figure \ref{region}).   
If X-ray and optical emissions come from the same region in B$\gamma$ scenario with non-dipole dominated case, 
X-ray luminosity is reduced by a factor of $\sim30$ because of the frequency dependence of the lower limit 
on the emission region $r_{\rm ct}\propto\nu_{\rm obs}^{-1/3}$ (equation \ref{r_ct})
and the distance dependence of the luminosity $L_{\rm syn}\propto r^{-3/2}$ (equation \ref{epsilon_syn,Bgamma}). 
Then, the efficiency $\eta$ should be an order of unity at least for old pulsars which the non-dipole magnetic field is 
required for the observed optical luminosity. 

\subsection{$\gamma$-ray Flux Ratio}
\label{sec:fluxdiscussion}

Since outer accelerator models such as the outer gap 
\citep[e.g., ][]{TWC11, WR11, H13, VTHP15a, VTHP15b, VT15, VTM15, Pier15, Pier16} 
and the current sheet models \citep[e.g., ][]{BS10, Kal+12a, P13, KHK14, CPS16} could explain 
the observed $\gamma$-ray emission features, 
$\gamma\gamma$ scenario is expected to work in $\gamma$-ray pulsars. 
However, the observed $\gamma$-ray to X-ray flux ratios for a significant fraction of $\gamma$-ray pulsars 
are much smaller than that in $\gamma\gamma$ scenario (left panel of figure \ref{ratio}). 
For some low-$L_{\rm sd}$ $\gamma$-ray pulsars, the observed $\gamma$-ray to optical flux ratios are also 
smaller than that in $\gamma\gamma$ scenario (right panel of figure \ref{ratio}). 
In order to explain the observed flux ratios, some additional mechanisms are required.
Since the low-$L_{\rm sd}$ pulsars also show the modulation phenomena in radio \citep{WMJ07}, 
the additional mechanisms may be related to the origins of those phenomena. 
We consider some possibilities to resolve the flux ratio discrepancies. 

In our model, we neglect the thermal emission from entire surface of a neutron star 
as seed photons in $\gamma\gamma$ pair creation.
Using minimum cooling scenario \citep{PLPS04}, 
the effective temperature of the entire surface is $T_{\rm sur}\gtrsim10^6$ K for the pulsars with their age 
$\tau \lesssim 10^5$ yr, which corresponds to the range of spin-down luminosity of $\gamma$-ray pulsars, 
$L_{\rm sd}\sim10^{36}$ erg s$^{-1}$.
For the age $\tau\sim10^5$ yr, the temperature is $T_{\rm sur}\sim10^6$ K and the surface thermal luminosity is 
$L_{\rm sur}\sim10^{33}$ erg s$^{-1}$, which corresponds to $\sim10^{-3}$-$10^{-2} L_{\rm sd}$. 
Then, the number of the seed photons $N_{\rm sur}\propto L_{\rm sur}/T_{\rm sur}$ is a factor of $\sim$3-30 times larger 
than that in our fiducial case with $L_{\rm pc}\sim10^{-3}L_{\rm sd}$ and $T_{\rm pc}\sim10^{6.5}$ K.
The seed photon number is proportional to the optical depth $\tau_{\gamma\gamma}$ (equation \ref{tau_gammagamma}) and 
the optical depth is proportional to the synchrotron luminosity $L_{\rm syn}\propto\tau_{\gamma\gamma}$ 
(equations \ref{L_syn} and \ref{dotN_s}). 
Taking into account the thermal emission from the entire surface, 
the luminosity $L_{\rm syn}$ could increase and the luminosity ratio $L_{\rm cur}/L_{\rm syn}$ 
could decrease by a factor of $\sim3$-30.
However, for pulsars with $L_{\rm sd}\lesssim10^{35}$ erg s$^{-1}$, 
the thermal photons from the entire surface do not contribute to the optical depth $\tau_{\gamma\gamma}$ 
\citep[e.g., ][]{YP04}.

We consider a possible solution that not only the outer accelerator works to emit 
the observed $\gamma$-ray emission, but also B$\gamma$ scenario could work to contribute to
the observed non-thermal emission to explain the observed X-ray and optical emission 
for low-$L_{\rm sd}$ $\gamma$-ray pulsars. 
Since the strong absorption feature have not been seen in the observed $\gamma$-ray spectra 
even in low-$L_{\rm sd}$ $\gamma$-ray pulsars \citep{Ab13}, 
the observed $\gamma$-ray emission comes from the outer magnetosphere where $\gamma\gamma$ scenario works. 
On the other hand, 
in $\gamma\gamma$ scenario, the observed luminosities in X-ray and optical bands exceed the maximum luminosity, 
and the observed $\gamma$-ray to X-ray and optical flux ratios $F_{\gamma}/F_{\rm X}$ and $F_{\gamma}/F_{\rm opt}$ 
are much smaller than the ratio $L_{\rm cur}/L_{\rm syn}$ for low-$L_{\rm sd}$ $\gamma$-ray pulsars.
In B$\gamma$ scenario, 
the observed X-ray and optical luminosities are lower than the maximum synchrotron luminosities. 
In addition, most of $\gamma$-ray photons are absorbed in B$\gamma$ process, 
so that the flux ratios $F_{\gamma}/F_{\rm X}$ and $F_{\gamma}/F_{\rm opt}$ could be reduced.
Then, if the most of the Poynting flux converts to the particle energy at the inner magnetosphere and the remaining
Poynting flux converts at the outer magnetosphere, the observed flux ratios should become lower than those 
in $\gamma\gamma$ scenario.
The coexistence of the multiple acceleration regions in the magnetosphere has been suggested by \citet{YS12, P13}. 
Recently, such emission models are also suggested to explain the observed light curves of $\gamma$-ray pulsar 
PSR J1813-1246 in X-ray and $\gamma$-ray \citep{Mar14}. 
For the pulsars with $L_{\rm sd}<10^{35}$ erg s$^{-1}$, the fraction of the population with $\gamma$-ray detection 
significantly decreases \citep{Laf15}. 
Although the distance, the inclination angle, and the viewing angle affect the $\gamma$-ray detectability, 
the relative activities of inner and outer accelerators may also affect it. 
The trend in the $\gamma$-ray detectability reported by \citet{Laf15} is consistent with the picture that
the energy conversion efficiency $\eta$ at the inner accelerator is larger 
than that at the outer accelerator for low-$L_{\rm sd}$ pulsars, while most of Poynting flux converts to the
particle energy flux at the outer magnetosphere for high-$L_{\rm sd}$ pulsars. 
The multiple emission regions have some constraints such that electromagnetic cascade sites 
do not reside on the same magnetic field line in the magnetosphere \citep[e.g., ][]{YS12, KAT16}. 
These constraints could be reflected in the shape of observed light curves \citep[e.g., ][]{RY95}. .
The non-dipole field could also affect the observed light curves. 
In particular, for the non-thermal X-ray emission from PSR J1741-2041, 
we should take into account the effect of non-dipole magnetic field at the synchrotron emission region 
(figure \ref{efficiency}). 
Unfortunately, the light curves of non-thermal component for most of low-$L_{\rm sd}$ pulsars are crude 
in current observations. 
We expect that further observations by such as {\it NuSTAR} and {\it NICER} 
will clarify the detailed shape of the light curve. 

There may be another energy source such as the dissipation of the stellar magnetic field 
as considered in the magnetar model \citep[e.g., ][]{DT92}. 
In our model, the total energy flux of primary particles is limited by the spin-down luminosity 
(equation \ref{dotN_p}). 
Additional energy sources could make the luminosity exceed the spin-down limit. 
In fact, 
magnetar-like bursts and following X-ray flux enhancements are detected from apparently rotation-powered pulsars, 
PSRs J1846-0258 and J1119-6127 \citep{Gav08, AKTS16}.
For PSR J1119-6127, the enhancement of the power-law component was seen in the X-ray spectrum \citep{AKTS16}, 
which emission mechanism may be the synchrotron radiation. 
Significant fraction of rotation-powered pulsars may also have the dissipative magnetic field, 
although bursting pulsars have relatively high magnetic field strength $B_{\rm s}\gtrsim B_{\rm q}$.
Here, we introduce the energy flux $L_{\rm B}$ via the dissipation of the magnetic field 
and its normalized value, $\epsilon_{\rm B}\equiv L_{\rm B}/L_{\rm sd}$.
We assume that the available magnetic field energy is comparable to the dipole field, 
$E_{\rm B}\sim(4\pi/3)R_{\rm ns}^3(B_{\rm s}^2/8\pi)$. 
Then, the condition for not exhausting the total magnetic energy $E_{\rm B}$ via the dissipation 
within their lifetime $\tau\sim\tau_{\rm c}$ is given by \citep[e.g., ][]{ZH00b}
\begin{eqnarray}\label{epsilon_B}
\epsilon_{\rm B}<E_{\rm B}/(L_{\rm sd}\tau).
\end{eqnarray}
From equation (\ref{epsilon_B}), the steady emission condition in the lifetime is derived as,
\begin{eqnarray}\label{steady-condition}
\dot{P}\gtrsim5\times10^{-11}\epsilon_{\rm B}P_0^{-3}~{\rm s~s}^{-1}.
\end{eqnarray} 
Assuming that most of the energy flux of dissipated magnetic field converts 
to the kinetic energy flux of the primary particle, 
the energy flux $\epsilon_{\rm B}$ has to be larger than unity to exceed the spin-down luminosity as an energy source. 
Only soft $\gamma$-ray repeaters and anomalous X-ray pulsars satisfy this condition \citep{ZH00b}. 
Even if we assume that all dissipated energy flux finally converts to the non-thermal luminosity, 
the parameter $\epsilon_{\rm B}$ should be an order of $\sim10^{-3}$ for X-ray band and $\sim10^{-5}$ for optical band
to cover the observed luminosity. 
We show lines of equation (\ref{steady-condition}) with $\epsilon_{\rm B}=10^{-3}$ and $10^{-5}$ 
in the left and right panels of figure \ref{p-pdot} as green dotted lines, respectively.
In X-ray band, no $\gamma$-ray pulsar satisfies the required condition with $\epsilon_{\rm B}=10^{-3}$. 
In optical band, an half of $\gamma$-ray pulsars could satisfy the condition with $\epsilon_{\rm B}=10^{-5}$.
Therefore, if the magnetic field dissipation is the dominant energy source, their non-thermal luminosity does not 
stay constant in their lifetime unless the dissipation of the non-dipole component 
dominates the total energy $E_{\rm B}$. 
The effects of the additional energy source should also be discussed in a model of magnetar wind nebula \citep{T16}.

There may be another mechanism to give the pitch angle to the secondary particles 
such as a plasma instability \citep[e.g., ][]{MU79, MLMV00} 
and cycrotron resonant absorption of radio photons \citep[e.g., ][]{LP98, HSDF08}. 
Then, a part of the momentum of secondary particles parallel to the magnetic field could be converted 
to the perpendicular component, and is used to emit via synchrotron radiation. 
This case corresponds to $\gamma\gamma$ scenario with $\alpha\sim1$.
Even in this case and assuming $\gamma_{\rm s,pair}\sim\gamma_{\rm s,syn}$, 
from equation (\ref{epsilon_syn,gamma-gamma}), 
the maximum efficiency $\epsilon_{\rm syn}$ is limited by the number of created secondary particles, 
$\epsilon_{\rm syn}\lesssim\tau_{\gamma\gamma}$. 
For typical $\gamma$-ray pulsars with $\epsilon_{\rm pc}\sim10^{-3}$, $T_{\rm pc}\sim10^6$ K and $P_0\sim0.1$s, 
the optical depth is $\tau_{\gamma\gamma}\lesssim10^{-4}$, 
which is smaller than the observed non-thermal efficiency in the X-ray band. 

The synchrotron radiation from pairs created by ingoing curvature photons may also contribute 
to the observed non-thermal emission.
We only consider that the primary particles emit curvature radiation with outgoing direction.
For the inwardly emitting curvature photons, the collision angle with thermal X-ray photons could be $\cos\theta\sim-1$. 
Then, the optical depth $\tau_{\gamma\gamma}$ could be $\sim (R_{\rm lc}/r)$ times higher than the outward case.
However, using the limitation of the luminosity of the thermal radiation from the heated polar cap, 
the conversion efficiency should be $\eta\lesssim10^{-2}$ even if we consider the cooling via curvature radiation 
for the energy flux of the ingoing primary particles. 
The expected flux of synchrotron radiation from pairs created from curvature photons emitted 
by ingoing primary particles is comparable to or lower than that from the outgoing pairs 
\citep{KT14, KT15}. 
Note that in some geometrical conditions, 
the inward synchrotron emission may dominantly contribute to the observed non-thermal emission 
\citep{TCS08, KK11, WTC13}.
While the observed non-thermal X-ray luminosity is within our model upper limit, 
since the inward $\gamma$-ray could be effectively absorbed, the flux ratio $F_{\rm X}/F_{\gamma}$ could be lower than 
our model lower limit.
The detailed light curve in both $\gamma$-ray and X-ray bands helps to show whether
the contribution of the inward emission to the observed non-thermal emission is less significant.

Except for the synchrotron radiation, 
inverse Compton scattering may work to contribute to the observed non-thermal emission 
\citep[e.g., ][]{ZH00}. 
For the non-resonant Compton scattering, 
the efficiency should be low because of the small value of the optical depth
as seen in $\gamma\gamma$ scenario. 
In the case of resonant Compton scattering, the resonant condition is 
\begin{eqnarray}\label{resonance}
E_{\rm pc}\gamma_{\rm s}(1-\cos\theta_{\rm col})=\hbar\frac{eB}{m_{\rm e}c},
\end{eqnarray}
where $\gamma_{\rm s}$ is the Lorentz factor of the secondary particles.
The energy of the scattered photons is
\begin{eqnarray}\label{scattered}
h\nu_{\rm obs}=2\gamma_{\rm s}m_{\rm e}c^2\frac{B}{B_{\rm q}}.
\end{eqnarray}
Using equations (\ref{resonance}) and (\ref{scattered}), 
the emission region in the case of outgoing primary particles is derived as
\begin{eqnarray}
r_{\rm res,6}\sim12\nu_{\rm obs,keV}^{-1/7}T_{\rm pc,6.5}^{-1/7}B_{\rm s,12}^{5/14}L_{\rm sd,31}^{-1/28}.
\end{eqnarray}
This is much smaller than the lower limit $r_{\gamma\gamma}$ 
from the condition of the $\gamma\gamma$ pair creation (equation \ref{r_gammagamma}). 
Only the inner accelerator where B$\gamma$ process works 
could supply the scattering particles at the resonant region $r_{\rm res}$.
Then, if the resonant Compton scattering contributes to the observed luminosity for the $\gamma$-ray pulsars, 
both the inner and outer accelerators work to produce $\gamma$-ray, X-ray and optical emissions.
Thus, the inverse Compton model also requires the existence of the multiple particle acceleration sites
in the magnetosphere. 

\acknowledgments
We are grateful to the anonymous referee for useful suggestions. 
We would  like to thank K. Asano, Y. Ohira, S. Shibata and J. Takata for fruitful discussions. 
This work is supported by KAKENHI 16J06773 (S.K.), 24000004 (S.J.T.).

\end{document}